\documentclass[aps,prl,preprint]{revtex4-2}

\usepackage{booktabs}   
\usepackage{mhchem}     
\usepackage{graphicx}   
\usepackage{subcaption} 
\usepackage{overpic}    

\begin{document}

\title{GPT-assisted learning of structure-property relationships by graph neural networks: Application to rare-earth doped phosphors}

\author{Xiang Zhang}
\author{Zichun Zhou}
\author{Chen Ming}
\author{Yi-Yang Sun}
\email{Email: yysun@mail.sic.ac.cn}
\affiliation{State Key Laboratory of High Performance Ceramics and Superfine Microstructure, Shanghai Institute of Ceramics, Chinese Academy of Sciences, Shanghai 201899, China}

\date{Oct. 10, 2023}

\begin{abstract}
Applications of machine learning techniques in materials science are often based on two key ingredients, a set of empirical descriptors and a database of a particular material property of interest. The advent of graph neural networks, such as Crystal Graph Convolutional Neural Network (CGCNN), demonstrates the possibility of directly mapping the relationship between material structures and properties without employing empirical descriptors. Another exciting recent advancement is in large language models such as OpenAI's GPT-4, which demonstrates competency at reading comprehension tasks and holds great promise for accelerating the acquisition of databases on material properties. Here, we utilize the combination of GPT-4 and CGCNN to develop rare-earth doped phosphors for solid-state lighting. GPT-4 is applied to data-mine chemical formulas and emission wavelengths of 264 \ce{Eu^{2+}}-doped phosphors from 274 papers. A CGCNN model is trained on the acquired dataset, achieving a test $R^2$ of 0.77. The model is then used to screen over 40,000 inorganic materials to make predictions on the emission wavelengths. We also demonstrate the possibility of leveraging transfer learning to fine-tune a bandgap-predicting CGCNN model towards the prediction of phosphor emission wavelengths. The workflow requires minimal human supervision, little domain knowledge about phosphors, and is generalizable to other material properties. 
\end{abstract}

\maketitle
\newpage


Following the introduction of blue LEDs in the 1990s\cite{nakamura}, white LEDs, and in particular simple phosphor-converted white LEDs (pc-wLEDs)\cite{rjxie} , have made substantial inroads towards replacing traditional lighting technologies\cite{annurev}, thanks to their energy efficiency, environmental friendliness and long lifespan. Lanthanides (e.g. \ce{Eu^{2+}}) activated phosphors offer good optical tunability in addition to good photostability and color performance\cite{lanthanide-review}. Moderated by the strength of crystal fields and the nephelauxetic effect, the 5d-4f optical transitions of \ce{Eu^{2+}}, one of the most common and efficient activators\cite{eu-common}, produce a tunable emission spectrum ranging from ultraviolet to deep red\cite{300}. In addition to trial-and-error, techniques including combinational chemistry, chemical unit substitution, and a single-particle diagnosis approach\cite{rjxie,data-driven} have been introduced for discovering new phosphor materials. The abundance of available data on the topic has also triggered recent studies employing various machine learning models\cite{ml-phosphor} such as regularized linear regression, multilayer perceptrons\cite{lasso-ann,too-close} and random forests\cite{rf}.

Graph neural networks\cite{gnn}, such as the Crystal Graph Convolutional Neural Network (CGCNN)\cite{cgcnn}, recently found applications in materials science\cite{gnn-review}. Their internal data structure of graphs, vertices and edges is well suited for the representation of crystals, atoms and bonds. Trained on databases such as the ICSD\cite{icsd}, Materials Project\cite{matproj},  and OQMD\cite{oqmd}, and combined with spectral\cite{spectral-conv} and real-space\cite{gcnn} convolutions on graphs, graph neural networks, unified under the conceptual framework of message-passing neural networks\cite{mpnn}, have demonstrated high accuracy\cite{gnn-benchmark} - sometimes comparable to first-principles calculations when benchmarked against experimental data\cite{cgcnn} - on compositionally diverse datasets. Outside of materials informatics databases, though, sourcing data in bulk remains a chore, with large volumes of academic journals largely untapped.

There has long been interest\cite{ceder-review-1,ceder-review-2} in extracting structural information from the unstructured text of academic publications using natural language processing (NLP) techniques. Efforts\cite{ceder,zeolite} and mainline tools such as ChemDataExtractor\cite{chemdataextractor}, OSCAR4\cite{oscar4} and ChemTagger\cite{chemtagger} employed a combination of heuristic rule-based approaches and specialized machine-learning NLP models. As the NLP field shifted towards large language models (LLMs)\cite{nlp-to-llm}, interests in applying LLMs to data-mining papers have emerged\cite{gpt-extract-1, bioxiv}. LLMs have revolutionized the NLP field ever since the introduction of OpenAI's original generative pre-trained transformer (GPT)\cite{gpt1} and Google's Bidirectional Encoder Representations from Transformers\cite{bert} in 2018. Despite being task-agnostic, large-scale generative deep-learning models with hundreds of billions of parameters, trained on large corpus of texts, can outperform even specialized state-of-the-art fine-tuned NLP models in the few-shot setting\cite{gpt3}. Starting from an attention-based transformer architecture\cite{attention,transformer} in place of more traditional recurrent neural networks like LSTMs\cite{lstm1997} and utilizing a two-stage training procedure consisting of generative pre-training and fine-tuning\cite{gpt1}, OpenAI's GPT-n series has seen several iterations\cite{gpt2,gpt3,gpt4}. The latest closed-source GPT-4 model, which incorporates Reinforcement Learning From Human Feedback\cite{rlhf-icml}, performs better than the average human at reading comprehension tasks\cite{gpt4}. 

\begin{figure}
    \centering
    \begin{subcaptiongroup}
        \subcaptionlistentry{}
        \label{fig:1-1}
        \begin{overpic}[width=.9\textwidth]{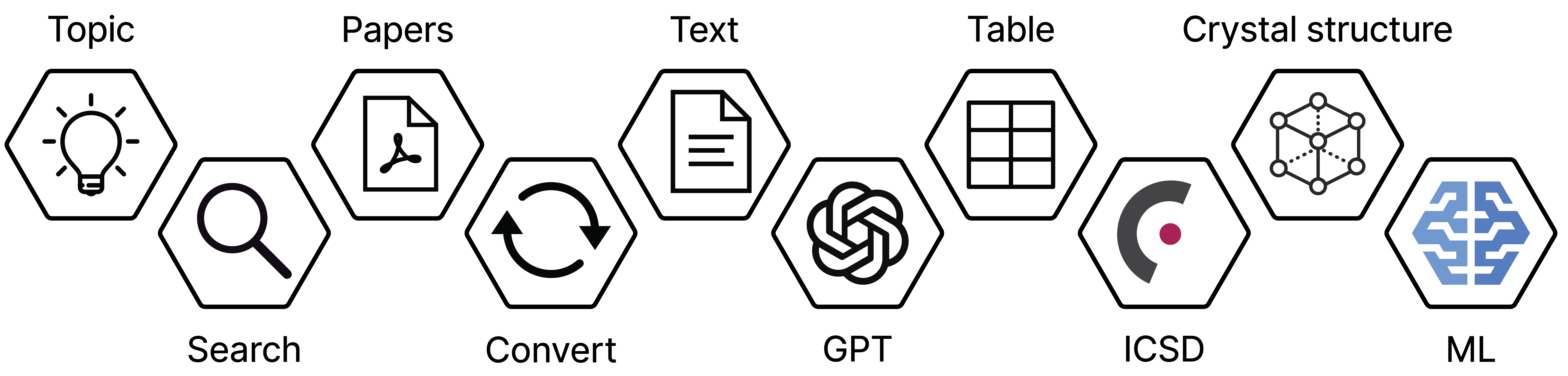}
            \put(-60,23) {\captiontext{}}
        \end{overpic}
        
        \vspace{3em}    
        
        \subcaptionlistentry{}
        \label{fig:1-2}
        \begin{overpic}[width=.9\textwidth]{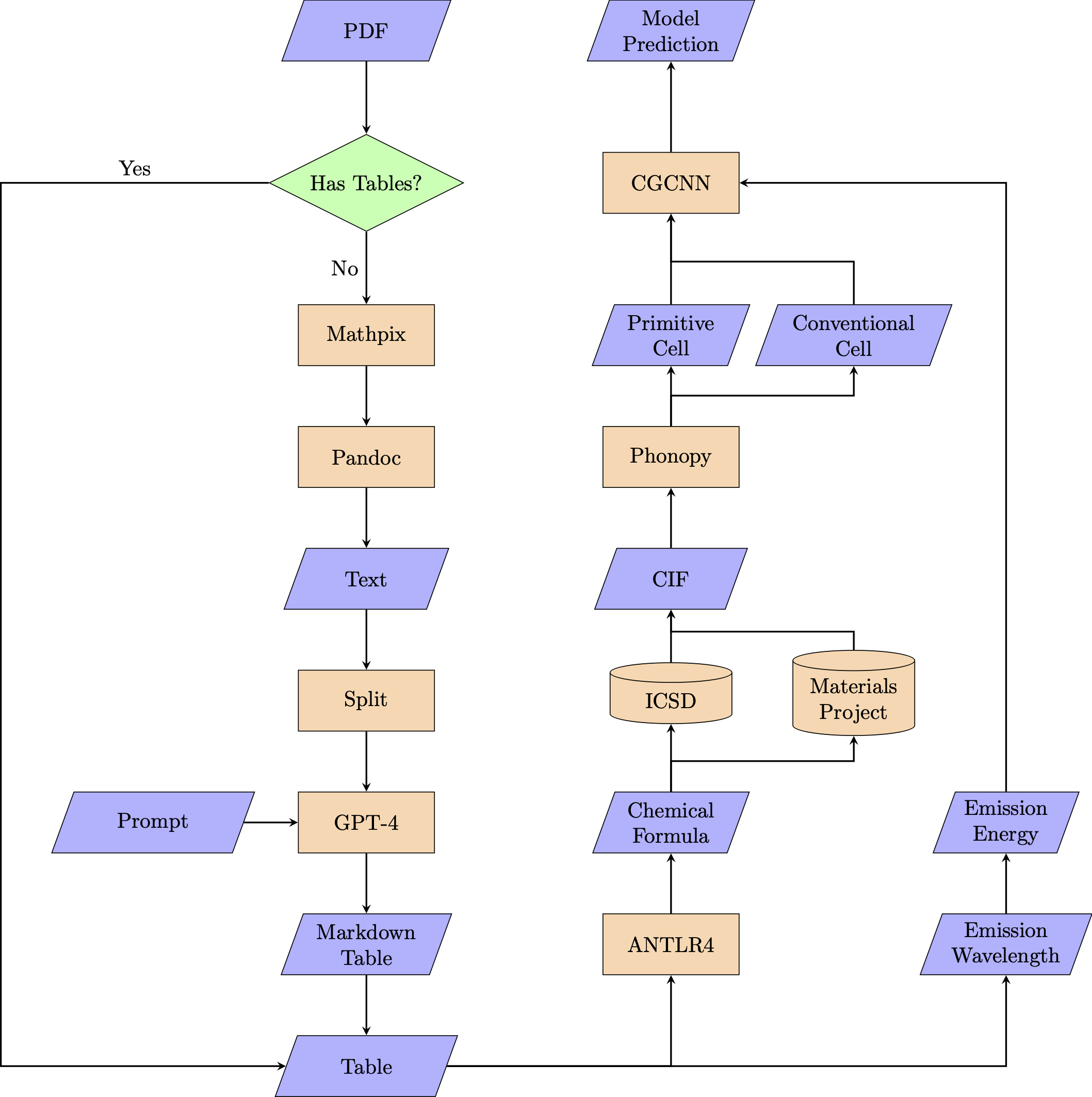}
            \put(-60,100) {\captiontext{}}
        \end{overpic}
    \end{subcaptiongroup}
    \captionsetup{subrefformat=parens}
    \caption{Data extraction and neural network training pipeline. \subref{fig:1-1} Overview of the pipeline in terms of data (upper hexagons) and their transformations (lower hexagons). \subref{fig:1-2} Flowchart with implementation details, as explained in the main text.}
\end{figure}

In this study, we present a data pipeline where we extract information about \ce{Eu^{2+}}-doped phosphors from scientific publications using GPT-4\cite{gpt4}, and train a CGCNN\cite{cgcnn} neural network using the extracted data (Fig. \ref{fig:1-1}). The pipeline requires minimal human supervision and depends on neural networks for both data acquisition and learning. Fig. \ref{fig:1-2} illustrates the pipeline in more implementation detail, including software tools and the logistical aspects of handling data. Starting from a collection of papers on \ce{Eu^{2+}}-doped phosphors, we transformed the papers in PDF format into text, used GPT-4 to parse the text into a table composed of chemical formulas and emission wavelengths, matched the chemical formulas to crystal structures in the ICSD database (in CIF format), and trained a CGCNN neural network. 

\begin{figure}[h]
    \centering
    \includegraphics[width=0.8\textwidth]{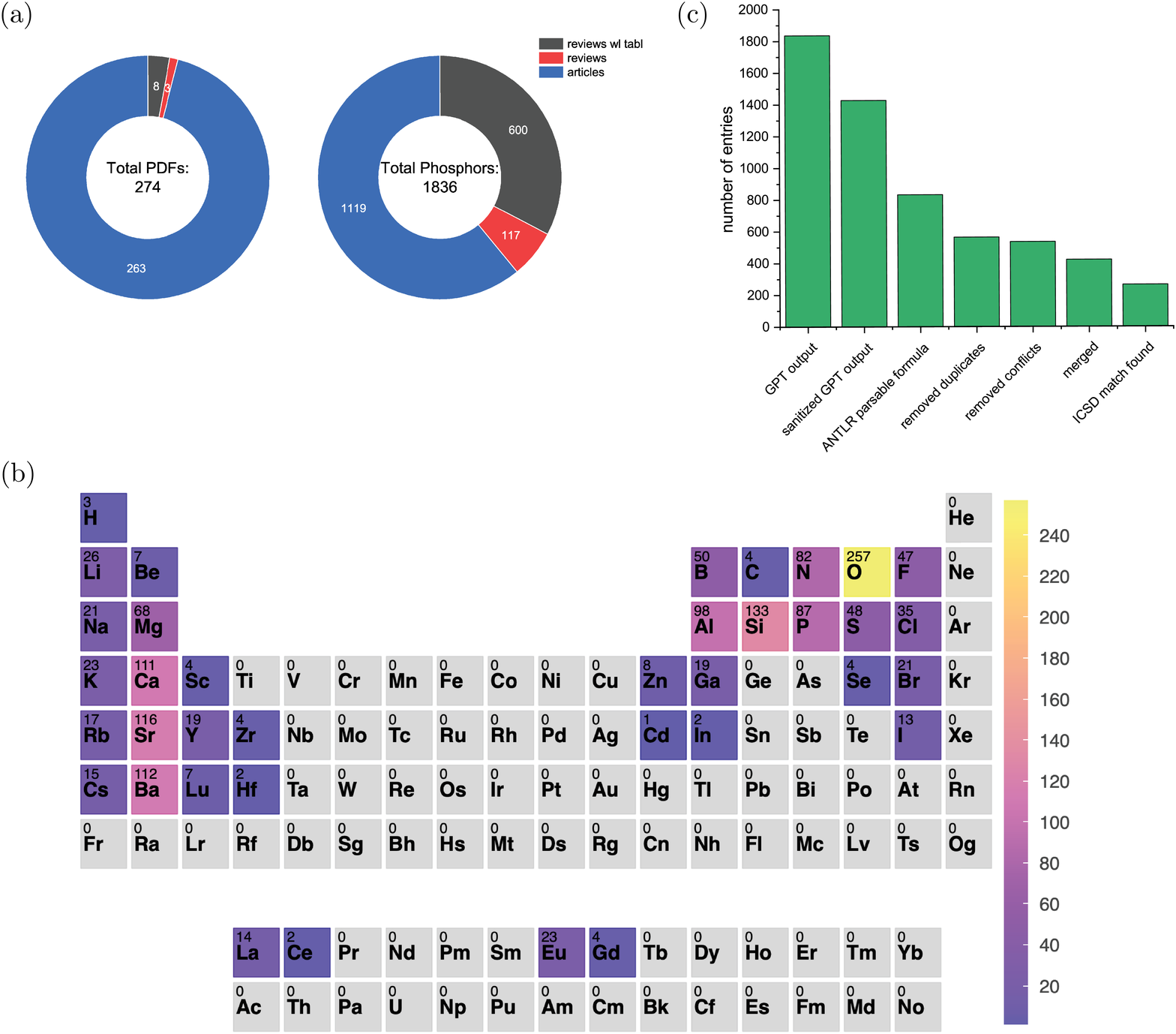}
    \caption{Descriptive statistics of the dataset. (a) Number of PDFs and GPT-4-parsed phosphor entries contributed by review and non-review papers. Review papers are further divided into two classes with and without tabulated dataset. (b) Periodic table showing the occurences of each element in host materials in the refined dataset.  (c) Number of phosphor entries in the dataset at each refining stage in the workflow.}
    \label{fig:2}
\end{figure}

We manually assembled a dataset comprising 274 papers (Table S1) on the topic of \ce{Eu^{2+}}-doped phosphors for data-mining.
We focused exclusively on \ce{Eu^{2+}}-doped phosphors, where data is most readily available, and did not include \ce{Ce^{3+}} or \ce{Eu^{3+}}-doped phosphors, to prevent the need for neural networks to generalize across different structure-property relationships for different activators. The 274-paper dataset consists of 11 reviews and 263 non-review papers (Fig. \ref{fig:2}a). While few in number, the reviews offer a more densely packed source of information, and in the case of tabulated data that would require no parsing using GPT-4, provide a trusted baseline against which the GPT-parsed data can be benchmarked against. Both the review and non-review papers contribute a significant portion to our final database. Fig. \ref{fig:2}b summarizes the chemical compositions of host materials, which include silicates, nitrides and oxynitrides, phosphates, aluminates, halides etc\cite{300,annurev}.

Due to the inherent messiness\cite{messy} of real-world data, of the over 1800 entries output by GPT-4, only 264 made it into the CGCNN training set. As shown in Fig. \ref{fig:2}c, the entries are discarded each stage down the pipeline. Several factors give rise to invalid data. Converting, or rather reconstructing text from PDFs, involve heuristics and inference, which introduces errors. Large language models may make mistakes when parsing text. Duplicate entries and inconsistencies between different literature sources exist, either due to errors within the pipeline or external errors. CIFs corresponding to specific chemical formulas might not be found in the ICSD. CGCNN does not understand CIFs with partial occupancies. Entries that could not be parsed into subsequent stages are discarded from the data pipeline.

PDF files were converted to LaTeX using Mathpix\cite{mathpix} (Fig. \ref{fig:1-2}), a commercial tool that demonstrated performance on par with state-of-the-art math text recognition methods\cite{mathpix-benchmark}, and LaTeX to plain text using Pandoc\cite{pandoc}, in preparation for parsing by GPT-4. For review papers containing summary data in tables, information was extracted by directly copy-pasting the tables. Alternatively, text could be scraped from HTML websites using heuristic rules\cite{boilerplate}, similar to using a browser's reader mode. We also evaluated the possibility of using Bing Chat (powered by GPT-4\cite{gpt4}) and Google Bard (powered by LaMDA\cite{lambda} and PaLM\cite{palm}), for aggregating web search results, as well as other PDF-to-text solutions. However, these results were unsatisfactory.


We next utilized OpenAI's GPT-4\cite{gpt4} and GPT-3.5\cite{gpt3} models via the Chat Completion API endpoint to extract information about phosphors, namely chemical composition and emission wavelength at the peak of luminescence spectrum, from the previously obtained bodies of text(Fig. \ref{fig:1-2}, Table S2). Since those models had demonstrated strong performance in reading-comprehension tasks, we anticipated that they would be effective at the task of literature review.

The GPT-4 model is limited in its context size, or the number of words it can process in one request, which makes it necessary to split the text to be analyzed into chunks (Fig. \ref{fig:1-2}). 
 Further, since we are mostly concerned with the small portion of each paper that mentions the emission wavelength in nanometers, we filtered out paragraphs that did not contain the word ``nm''.

\begin{figure}
    \centering
    \begin{subcaptiongroup}
        \subcaptionlistentry{}
        \label{fig:3-1}
        \begin{overpic}[width=.47\textwidth]{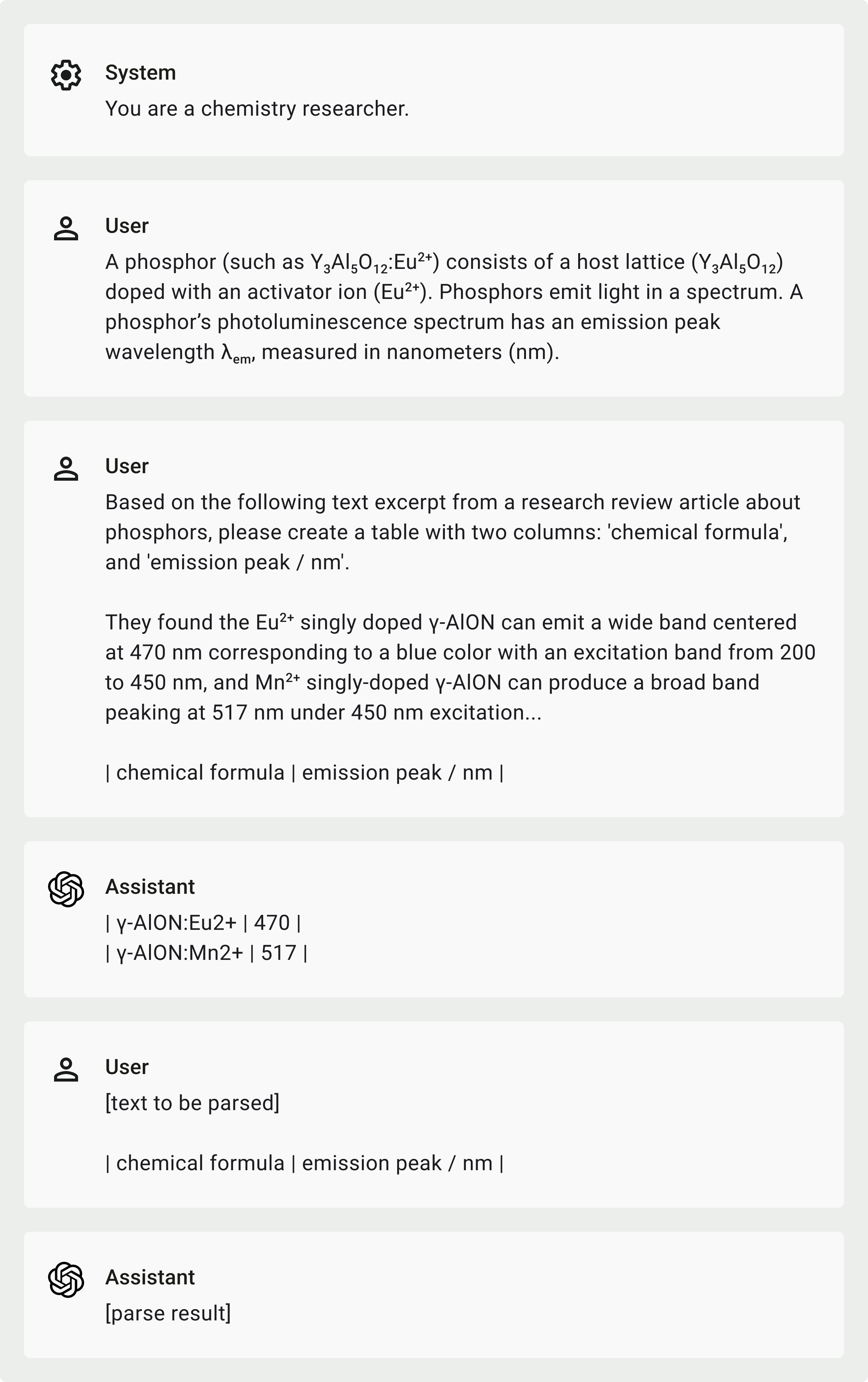}
            \put(-70.5,100) {\captiontext{}}
        \end{overpic}
    \end{subcaptiongroup}
    \hfill
    \begin{subcaptiongroup}
        \subcaptionlistentry{}
        \label{fig:3-2}
        \begin{overpic}[width=.47\textwidth]{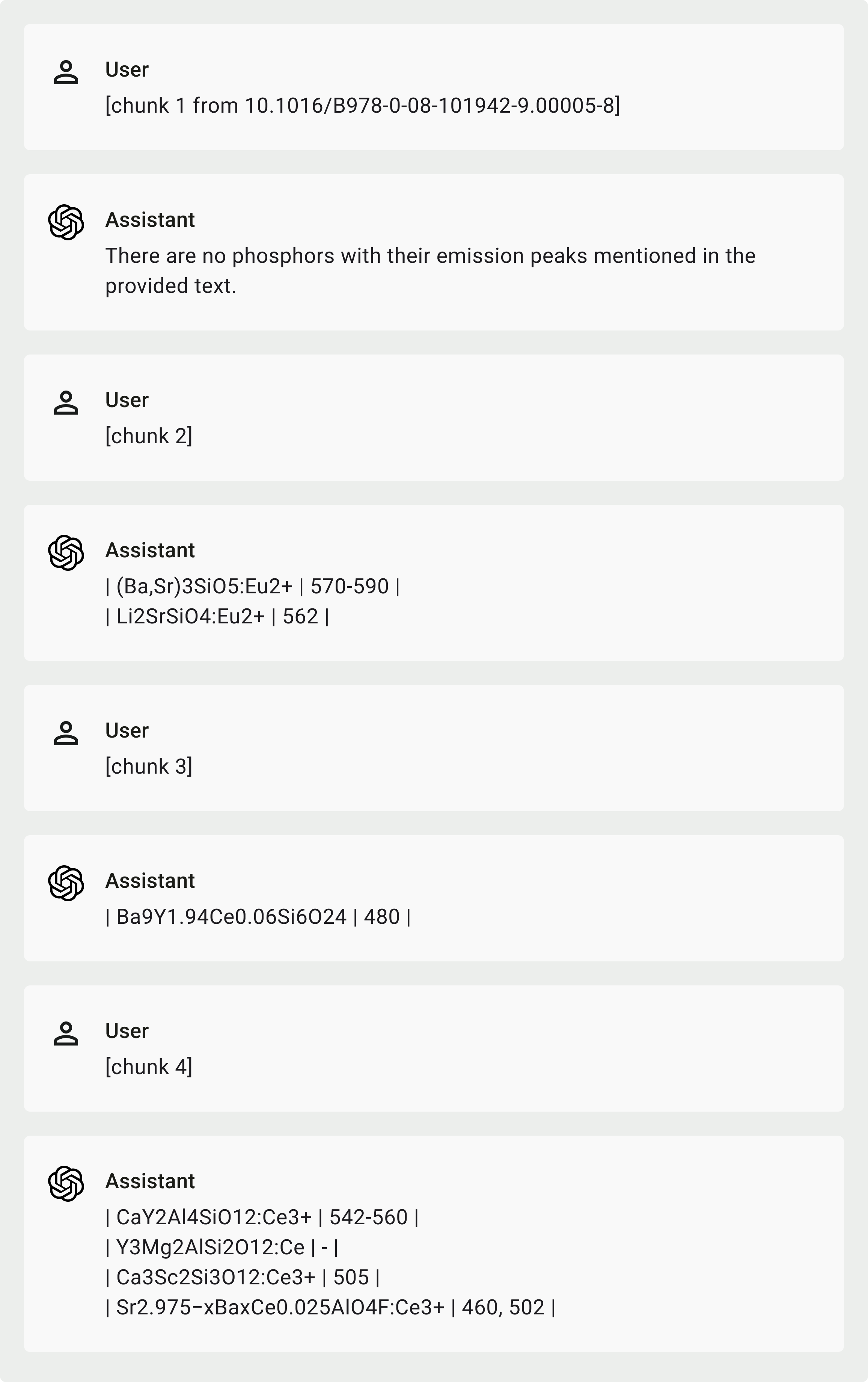}
            \put(-70.5,100) {\captiontext{}}
        \end{overpic}
    \end{subcaptiongroup}

    \vspace{3em}
    
    \begin{subcaptiongroup}
        \subcaptionlistentry{}
        \label{fig:3-3-1}
        \begin{overpic}[width=.31\textwidth]{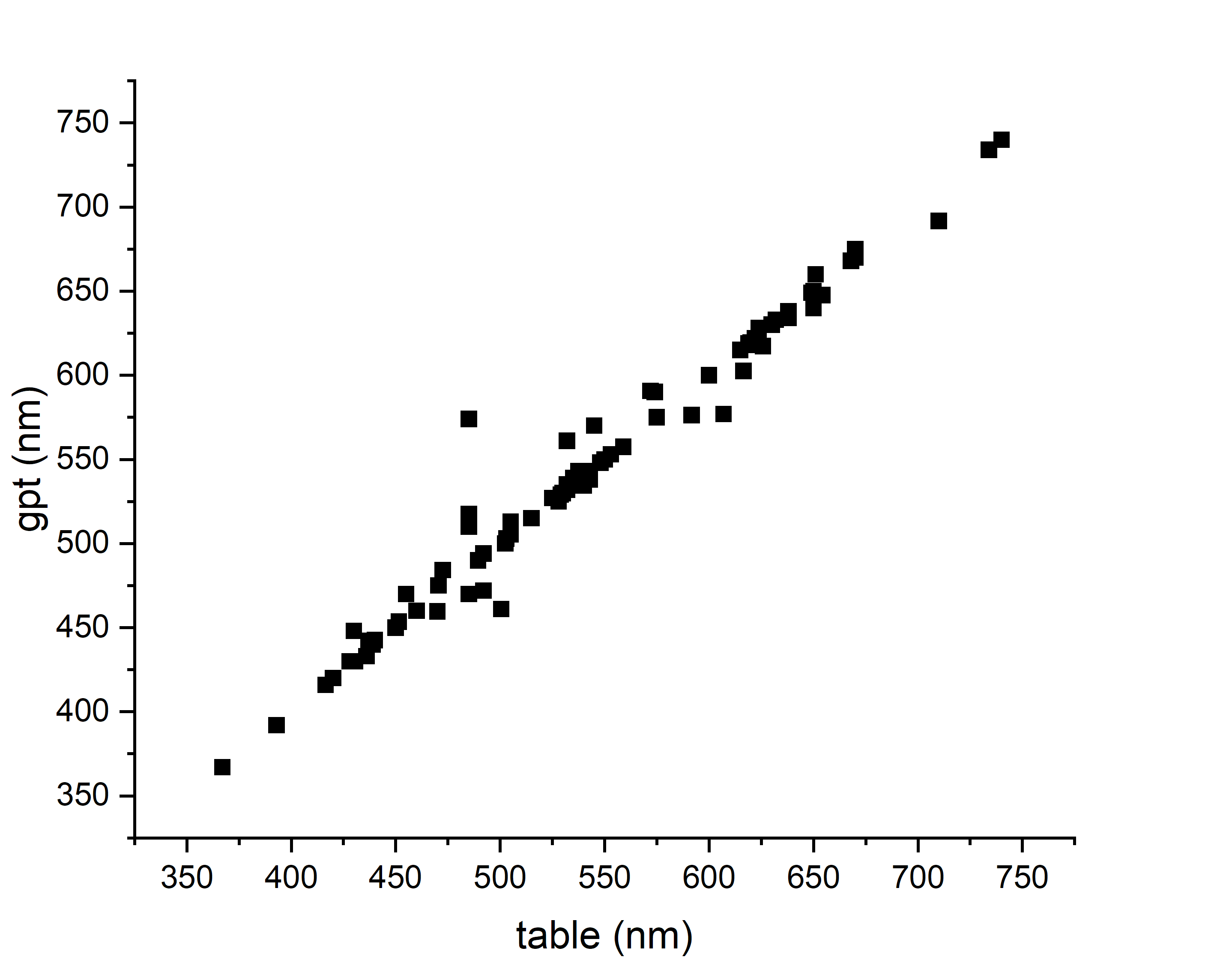}
            \put(-170,75) {\captiontext{}}
        \end{overpic}
        \hfill
        \subcaptionlistentry{}
        \label{fig:3-3-2}
        \begin{overpic}[width=.28\textwidth]{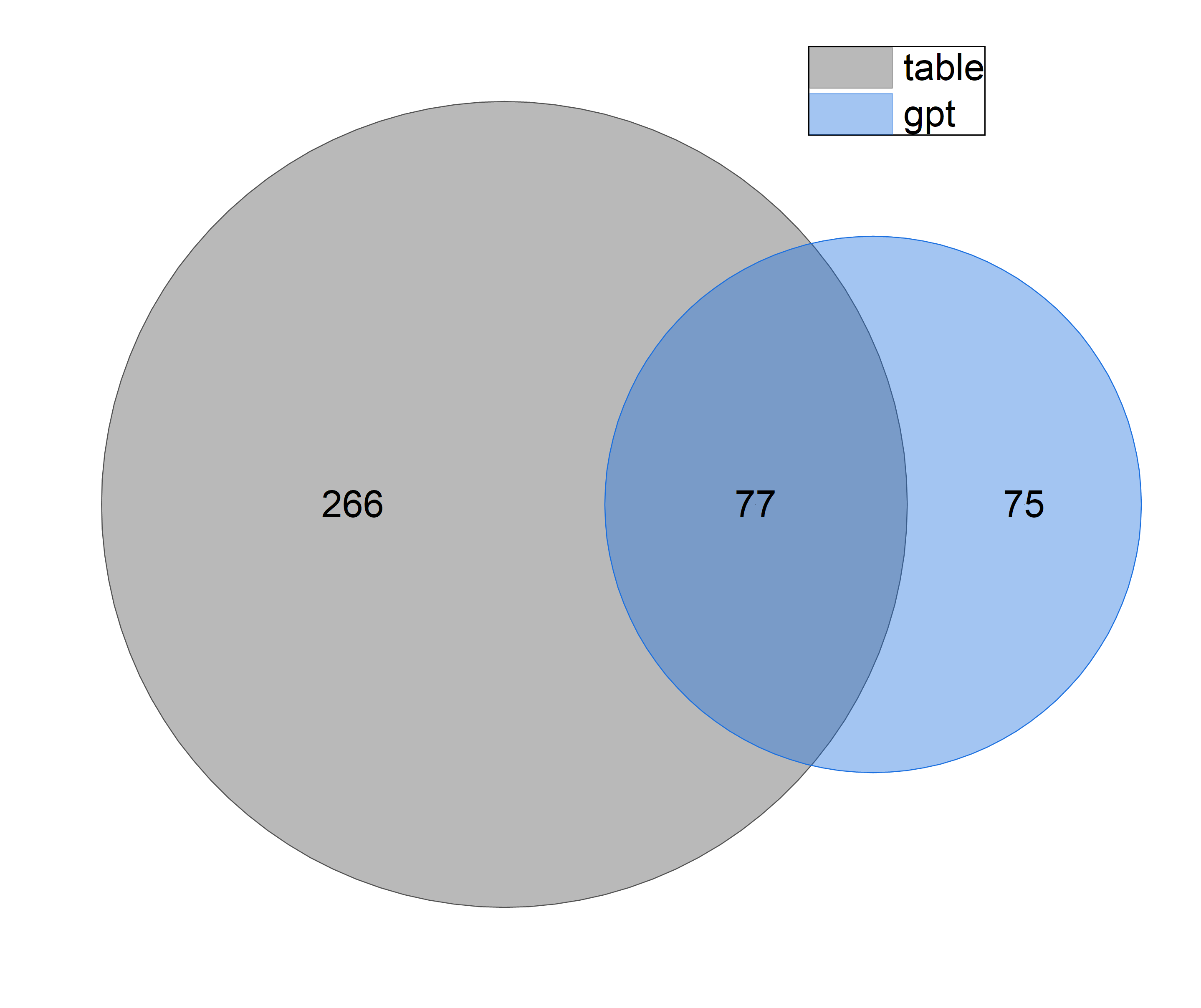}
            \put(-185,82) {\captiontext{}}
        \end{overpic}
        \hfill
        \subcaptionlistentry{}
        \label{fig:3-3-3}
        \begin{overpic}[width=.28\textwidth]{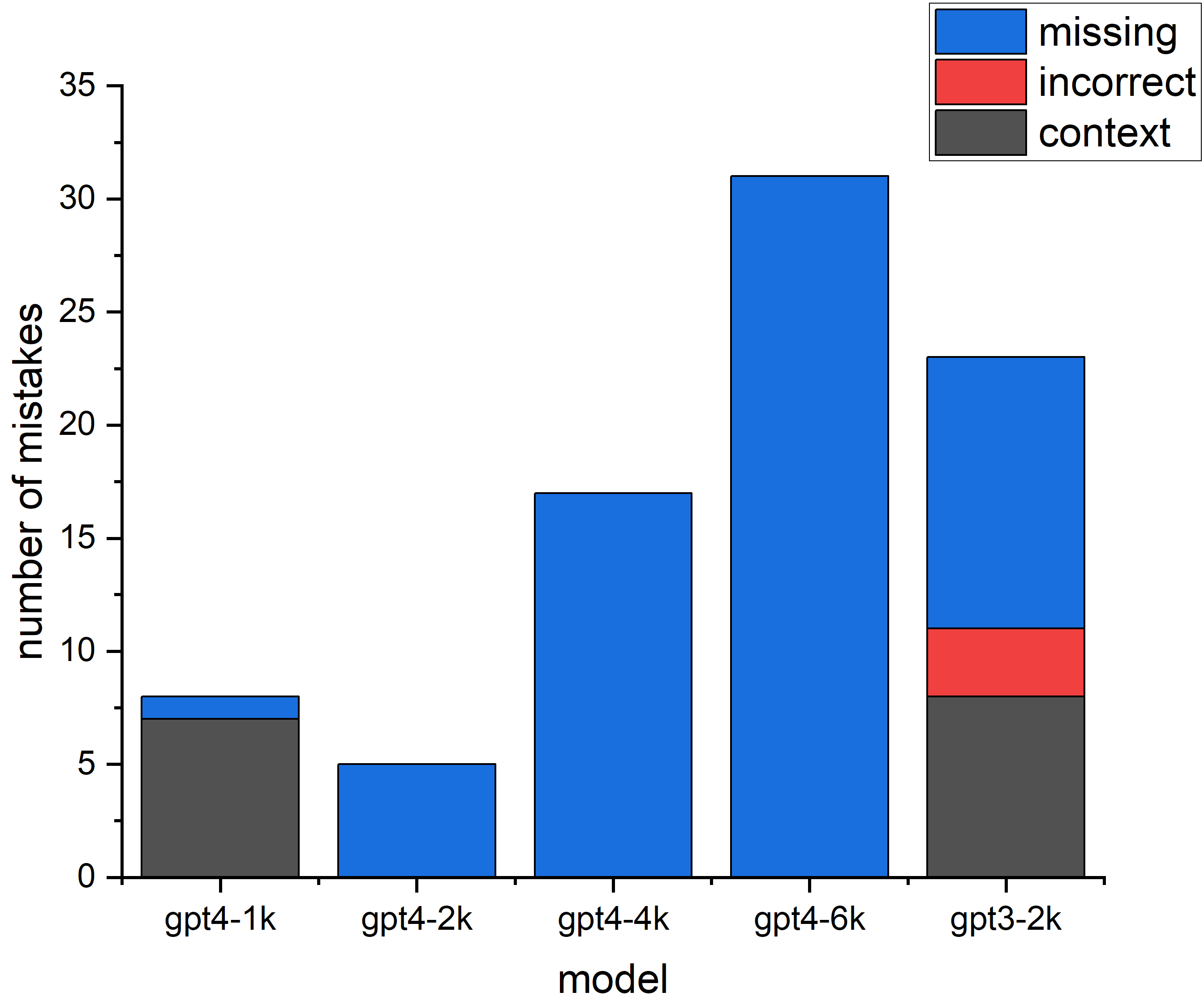}
            \put(-185,82) {\captiontext{}}
        \end{overpic}
    \end{subcaptiongroup}
    \captionsetup{subrefformat=parens}
    \caption{Prompt-engineering and benchmarking GPT-4. \subref{fig:3-1} Prompts instructing GPT-4 to extract chemical formulas and emission wavelengths, including a sample exchange for few-shot learning and ground truth regarding phosphors. \subref{fig:3-2} Sample raw GPT-4 output, illustrating challenges in data cleaning. \subref{fig:3-3-1} Comparison of GPT-4-extracted wavelengths against tabularized data from review articles. \subref{fig:3-3-2} Venn diagrams comparing database entries derived from tabular data in review articles against entries extracted by GPT-4. 
    \subref{fig:3-3-3} Analyzing GPT-4's inaccuracies. Models with longer memory understands context better, but forgets more. GPT-3.5 is inferior to GPT-4 and can make mistakes. }
    \label{fig:3}
\end{figure}

The task of instructing GPT-4 to extract the desired information from bodies of text falls into the category of prompt engineering\cite{gpt2}. Properly instructed, the task-agnostic GPT model can outperform state-of-the-art specialized NLP models\cite{gpt1}, such as those specializing in data-mining papers\cite{chemdataextractor,oscar4,chemtagger}. Fig. \ref{fig:3-1} outlines the prompt we used to tell GPT-4, in simple terms, to create a table of chemical formulas and their corresponding emission wavelengths, based on the given text.

We now explain the structure of and ideas behind our prompt (Table S5) in more detail. The GPT chat completion endpoint, similar to ChatGPT, takes as input a series of messages, a dialogue history consisting of interactions between the system (which designates how the chatbot should behave), the user, and the assistant (the chatbot itself). We designed our prompt based on the ``Parse Unstructed Data'' example from OpenAI's documentation. Based on the documentation's suggestion to provide examples, we included with every API call a fabricated sample exchange that demonstrates the model's desired behavior. Following the suggestion to ask GPT to adopt a persona, we asked GPT-4 to roleplay a chemistry researcher. While GPT-4 by itself already demonstrates general knowledge on STEM topics\cite{gpt-stem}, we provided more context on phosphors, chemical formulas, emission and absorption wavelengths, in the way they are commonly worded. HTTP requests are made using OpenAI's Python library, together with \verb|concurrent.future| for parallelization and \verb|tenacity| for retrying.

Fig. \ref{fig:3-2} shows a sample raw GPT-4 output generated by parsing a portion of Ref.{} \cite{annurev}. It has a few issues. GPT-4 always complains if there are no phosphors, and does not comply with instructions otherwise. The extracted chemical formulas do not always adhere to the host:dopant convention such as \ce{YAG{:}Ce^{3+}}. Formulas containing variables in them may require arithmetics, which GPT struggles to perform\cite{gpt-math}. Instead of attempting to fix each issue, we simply try to parse and sanitize each entry in GPT-4's output, and discard the entry if the parsing fails. We also tried asking GPT-4 to cleanup the table while giving it examples of the desired entry, but the improvements were insignificant. It is apparent that some entries, despite failing to be parsable in various creative ways, can be easily corrected and salvaged with a little human discretion, which we applied. 


We now evaluate and discuss GPT-4's performance in this data extraction task. The eight review papers with tables, which did not require GPT-4 to parse, serve as a benchmark for evaluating GPT-4's accuracy. Fig. \ref{fig:3-3-1} compares the emission wavelengths extracted by GPT-4 against those from review tables (Fig. \ref{fig:3-3-2}). For 74 out of 77 entries, GPT-parsed results differ from the values in the review tables by less than 30~nm. Thus, when handling duplicate and conflicting entries, we discard all entries where the standard deviation exceeds 30~nm. 

We manually reviewed GPT-4's performance on Ref.{} \cite{annurev} (Table S3). Some notable observations are described as follows. GPT-4 can understand context. For example, when parsing ``the \ce{M2MgSi2O7} compound is slightly different\ldots the Ca, Sr and Ba variants have emission peaks at approximately 475~nm, 540~nm, and 500~nm'', ``the Ca variants'' is correctly understood to be \ce{Ca2MgSi2O7}, although this behavior appears to be influenced by temperature and confidence settings. GPT-4 is able to distinguish between excitation and emission peaks. For instance, ``excitations of \ce{YAG{:}Ce}\ldots occurs in a broad range, with a maximum at 460~nm, and emission occurs at 540~nm'' is correctly understood, though GPT-4 does not translate \ce{YAG} to \ce{Y3Al5O12}. GPT-4 rephrases ``Ce-doped \ce{Ca3Sc2Si3O12}'' into \ce{Ca3Sc2Si3O12{:}Ce}, as instructed to in the examples provided. GPT-4 might occasionally miss entries, but does not hallucinate non-existing ones, or make blatant mistakes.

It may be possible to gain some insight into the GPT-4's performance by analyzing the mistakes it makes on Ref.{} \cite{annurev} (Fig. \ref{fig:3-3-3}). As we feed GPT-4 longer bodies of text, it gets slightly better at understanding context, making fewer context-related mistakes, but forgets more entries. We found it optimal to split long text into 2000-token-long chunks, despite the 8000-token upper limit, and applied this setting when parsing other review and non-review papers. It is worthwhile noting that the GPT-3.5 model is the only model that blatantly makes mistakes, ``forgets'' more entries at 2k tokens and is much worse at understanding context than the GPT-4 model. One might conclude that GPT-3.5 is unreliable, whereas GPT-4 appears to be more, perhaps even adequately, trustworthy.


We then merged the GPT-4-parsed chemical formulas and those directly taken from tabular data in reviews into a single table, parsed the chemical formulas using ANTLR\cite{antlr}, queried ICSD for matching CIFs, and prepared the training set for CGCNN (Fig. \ref{fig:1-2}). We used ANTLR4, a parser generator, to develop a custom library for parsing chemical formulas into Pythonic objects, using a formal grammar definition for chemical formulas like Ba[(Mg(0.2)Li(1.8))(Al(2.2)Si(1.8))N6]:Eu2+, the nested parenthesis being a side effect of interpreting arithmetic expression like \ce{Al_{2+x}Si_{2-x}}. Existing libraries fell short in various categories. The Atomic Simulation Environment\cite{ase} and Python Materials Genomics\cite{pymatgen} libraries refuse to handle square brackets, while python libraries like chemparse, chempy and regular expressions in general cannot handle nested parentheses. We discarded entries that did not conform to our grammar, those with non-integer stoichiometry, and those not doped with \ce{Eu^{2+}} or whose dopants were not mentioned. We also reduced the integer stoichiometries to their simplest forms to facilitate comparison and querying.

We then queried the ICSD databases for matching CIF files (Fig. \ref{fig:1-2}). In the occasional cases where multiple matches are found, which are mostly duplicates (Fig. S1, S2, S3), we selected the entry having the fewest atoms. We discarded CIFs with partial occupancies because CGCNN is unable to parse them. We also used the Phonopy\cite{phonopy} program to find the primitive cells of the crystal structures. We also converted emission wavelengths to emission energies (in eV) in preparation for CGCNN consumption. We neither performed relaxation nor doped the host lattice. 


Up to this point, we have converted plain text descriptions of \ce{Eu^{2+}}-doped phosphors to a table of chemical formulas and emission energies, to a mapping from CIFs to emission energies, which will be input to CGCNN neural networks (Fig. \ref{fig:1-2}) for learning. The CGCNN model translates crystal structures into graphs. Compared to manually engineered fingerprints of the local chemical environment\cite{behler-parrinello,too-close}, the graphs hopefully serve as a less arbitrary and more lossless representation. An embedding layer transforms a one-hot encoding of elements into a 128-dimensional embeddings space for each node, after which graph convolutional, pooling, fully connected and readout layers are applied.

The 264-entry dataset was shuffled and randomly divided into 60\% training, 20\% validation and 20\% test set. Each entry comprised a crystal structure in the form of a CIF file, sourced from the ICSD and converted into the primitive cell, and a label in the material's emission energies. The task was supervised regression. We trained our CGCNN model using the original implementation in PyTorch\cite{cgcnn,pytorch}. Mean-squared-error losses were minimized using the Adam optimizer\cite{adam}. Early stopping, or a normalization strategy similar to it, was performed by completing the full number of 200 epochs and selecting the model at the epoch when validation accuracy reached its peak before declining (Fig. S5). For each class of hypotheses, we reshuffled the training-validation-test sets and repeated the whole process 10 times.

\begin{figure}
    \centering
    \includegraphics[width=.9\textwidth]{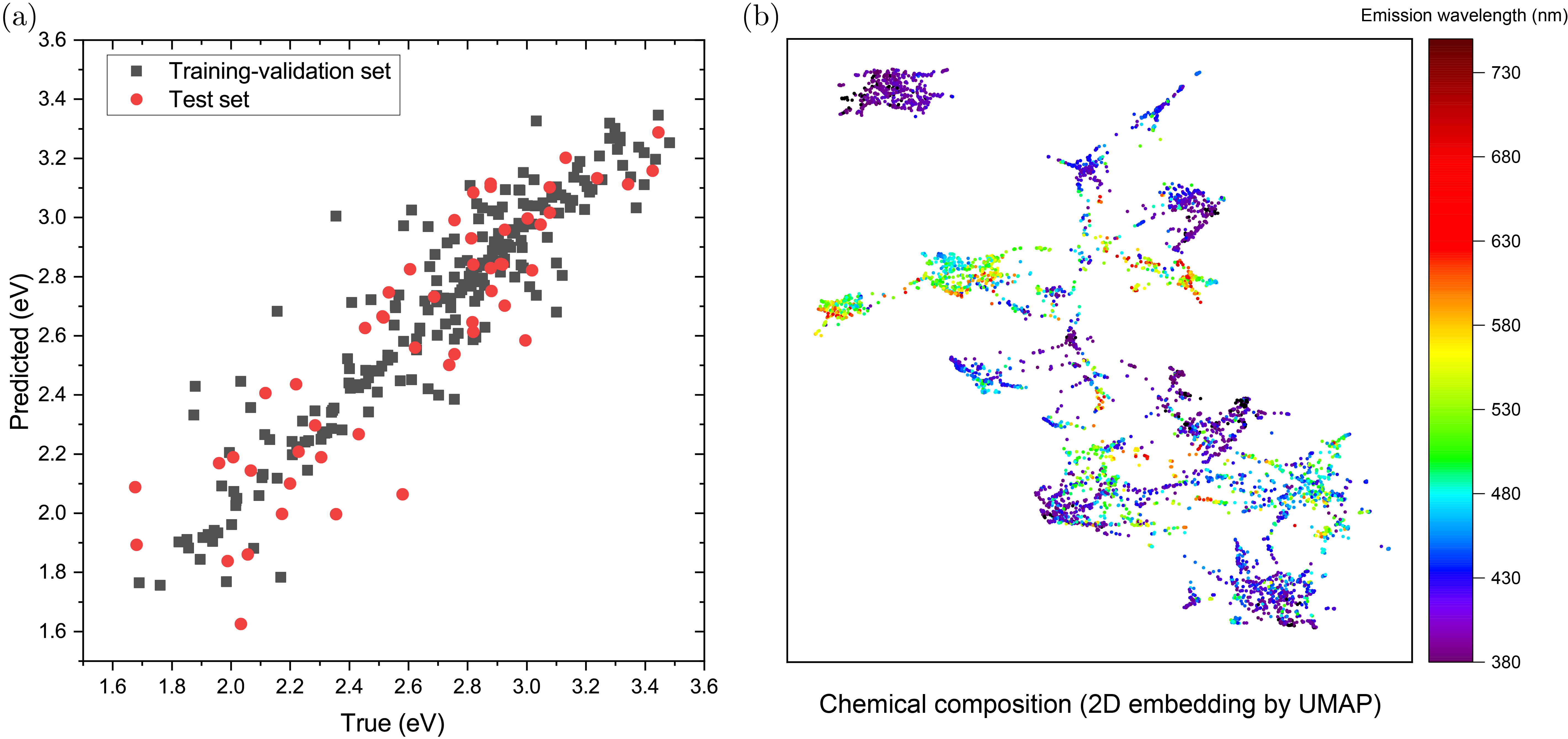}
    \caption{CGCNN performance and predictions. (a) Training-validation-test performance of CGCNN on the GPT-4-extracted dataset. $R^2$=0.77. (b) Using CGCNN to predict emission wavelengths across the ICSD dataset. Colors correspond approximately to emission wavelengths in the visible spectrum. X and y coordinates represent the chemical formulas through a UMAP embedding of one-hot encoding of atomic numbers.}
    \label{fig:4}
\end{figure}

A 10-fold (Table S4) average $R^2$ of 0.69 and MAE of 0.20 eV were achieved when training on primitive cells, obtained from the ICSD, with emission energies as labels and adopting the default embedding dimension of 64, 3 graph convolutional layers, a 128-wide fully connected and a readout layer. The best single model test case (Fig. \ref{fig:4}) yields an $R^2$ of 0.77 and MAE of 0.20 eV (see Table~\ref{fig:4-3}). 

\begin{table}
    \centering
    \caption{Comparison of CGCNN model variants, data and hyperparameter choices, and their impact on prediction accuracy for emission peak wavelengths (and energies) of \ce{Eu^{2+}} phosphors}
    \label{fig:4-3}
    \begin{tabular}{lccc}
        \toprule 
        \textbf{Model} & \textbf{R\textsuperscript{2}} & {\textbf{MAE / eV}} & \textbf{MAE / nm} \\ 
        \midrule 
        Highest single shot accuracy & 0.77 & 0.18 & -- \\
        Best model on average & 0.69 & 0.20 & -- \\
        \midrule
        Materials Project (instead of ICSD) & 0.68 & 0.19 & -- \\
        Conventional cell (instead of primitive cell) & 0.64 & 0.21 & -- \\
        Predict wavelengths (instead of energies) & 0.62 & -- & 39 \\
        SGD optimizer (instead of Adam optimizer) & 0.57 & 0.22 & -- \\
        Double hidden layer size & 0.54 & 0.22 & -- \\
        Transfer learning & 0.43 & 0.21 & -- \\
        \bottomrule 
    \end{tabular}
\end{table}

Table~\ref{fig:4-3} summarizes other data sources and architectures that we considered. We preferred energies to wavelengths as labels. One reason for doing so is that the dataset is already skewed towards the low-wavelength end, which can be alleviated by taking the inverse from wavelength to energy. We had also hoped that, by converting the ICSD CIFs to primitive cells, the unit cells would be smaller, making it easier for the model to focus on the relevant chemical environment of potential dopant sites. This seems to slightly improve the result as well. Sourcing CIFs from Materials Project\cite{matproj} instead of the ICSD yields more (276 instead of 264) entries and comparable $R^2$, although the results are more consistent across the 10 repetitions with a standard error of 0.02 instead of 0.06 eV. The Adam\cite{adam} optimizer outperforms the SGD optimizer, though not significantly, and not at all in some instances when training on wavelengths. Tuning the dimensions of the neural network produced no better results.

We then used the best model to predict the \ce{Eu^{2+}}-activated emission wavelengths of all inorganic compounds in the ICSD database (Fig. \ref{fig:4}b). To improve the quality of the prediction dataset, we removed duplicate entries, metal-organic frameworks and theoretical structures, and structures with partial occupancies, resulting in a 42284-entry dataset. Utilizing bootstrap aggregation, we predicted the emission wavelengths for the dataset using our top-performing models before taking the average. We first removed the materials containing elements that do not exist in the experimental dataset. We then assessed the reliability of our predictions based on the discrepancies between the outputs from different models. The assessment resulted in 5289 trusted, non-extrapolating predictions. 

To visualize the relationship between the chemical compositions of the 5289 materials and their predicted emission wavelengths, we used the Uniform Manifold Approximation and Projection (UMAP) method\cite{umap}, a dimensionality reduction technique, for embedding the chemical composition space into two dimensions. As a result, materials with similar chemical compositions end up near each other. In Fig. \ref{fig:4}b, the colors correspond to the predicted emission wavelengths. Materials of similar compositions indeed tend to emit light at similar wavelengths. The unsupervised UMAP plot reveals certain inner structures about classes of materials in the ICSD database that seem to be good indicators of emission light colors.

Last but not the least, we explored an important idea regarding the training of CGCNN models, namely, transfer learning. Xie et al.\cite{cgcnn} had trained their CGCNN models on the formation energies and bandgaps in the Materials Project database. One might naturally wonder, then, if some low-level physics governing the band structures of 100,000 inorganic compounds, perhaps the most basic notations of chemical bonds, would transfer over to the laws governing photoluminescenc. Our results suggested that it is possible. By continuing training from the bandgap model and freezing the embedding layer, we were able to transfer some generalizable knowledge in the form of neural network weights from the bandgap-predicting model and fine-tune it to predict phosphor emission energies instead, with minimal increase in MAE (Table~\ref{fig:4-3}).

In summary, we have demonstrated the ability of GPT-4 to efficiently and reliably extract information from research papers, and CGCNN to produce accurate and robust regression models ($R^2$=0.77) on a dataset of 264 rare-earth doped phosphors. We implemented a data pipeline incorporating GPT-4 and CGCNN, and trained a CGCNN model, which is used for predicting emission wavelengths of over 40,000 \ce{Eu^{2+}}-doped phosphors. In addition, applying the concept of transfer learning, we showed that a pre-trained CGCNN model, designed to predict bandgaps in the Materials Project database, can apply the same knowledge in the form of crystal structure embeddings to the prediction of phosphor emission wavelengths.

\section{Acknowledgements}
X.Z.\ is grateful to Professor J.\ C.\ Grossman for his guidance and the opportunity to study in his group.
This work is supported by the National Key Research and Development Program of China under Grant No. 2021YFB3500501.

\bibliography{references}

\begin{thebibliography}{57}%
\makeatletter
\providecommand \@ifxundefined [1]{%
 \@ifx{#1\undefined}
}%
\providecommand \@ifnum [1]{%
 \ifnum #1\expandafter \@firstoftwo
 \else \expandafter \@secondoftwo
 \fi
}%
\providecommand \@ifx [1]{%
 \ifx #1\expandafter \@firstoftwo
 \else \expandafter \@secondoftwo
 \fi
}%
\providecommand \natexlab [1]{#1}%
\providecommand \enquote  [1]{``#1''}%
\providecommand \bibnamefont  [1]{#1}%
\providecommand \bibfnamefont [1]{#1}%
\providecommand \citenamefont [1]{#1}%
\providecommand \href@noop [0]{\@secondoftwo}%
\providecommand \href [0]{\begingroup \@sanitize@url \@href}%
\providecommand \@href[1]{\@@startlink{#1}\@@href}%
\providecommand \@@href[1]{\endgroup#1\@@endlink}%
\providecommand \@sanitize@url [0]{\catcode `\\12\catcode `\$12\catcode
  `\&12\catcode `\#12\catcode `\^12\catcode `\_12\catcode `\%12\relax}%
\providecommand \@@startlink[1]{}%
\providecommand \@@endlink[0]{}%
\providecommand \url  [0]{\begingroup\@sanitize@url \@url }%
\providecommand \@url [1]{\endgroup\@href {#1}{\urlprefix }}%
\providecommand \urlprefix  [0]{URL }%
\providecommand \Eprint [0]{\href }%
\providecommand \doibase [0]{https://doi.org/}%
\providecommand \selectlanguage [0]{\@gobble}%
\providecommand \bibinfo  [0]{\@secondoftwo}%
\providecommand \bibfield  [0]{\@secondoftwo}%
\providecommand \translation [1]{[#1]}%
\providecommand \BibitemOpen [0]{}%
\providecommand \bibitemStop [0]{}%
\providecommand \bibitemNoStop [0]{.\EOS\space}%
\providecommand \EOS [0]{\spacefactor3000\relax}%
\providecommand \BibitemShut  [1]{\csname bibitem#1\endcsname}%
\let\auto@bib@innerbib\@empty
\bibitem [{\citenamefont {Nakamura}\ \emph {et~al.}(1994)\citenamefont
  {Nakamura}, \citenamefont {Mukai},\ and\ \citenamefont {Senoh}}]{nakamura}%
  \BibitemOpen
  \bibfield  {author} {\bibinfo {author} {\bibfnamefont {S.}~\bibnamefont
  {Nakamura}}, \bibinfo {author} {\bibfnamefont {T.}~\bibnamefont {Mukai}},\
  and\ \bibinfo {author} {\bibfnamefont {M.}~\bibnamefont {Senoh}},\ }\bibfield
   {title} {\bibinfo {title} {Candela-class high-brightness ingan/algan
  double-heterostructure blue-light-emitting diodes},\ }\href@noop {}
  {\bibfield  {journal} {\bibinfo  {journal} {Applied Physics Letters}\
  }\textbf {\bibinfo {volume} {64}},\ \bibinfo {pages} {1687} (\bibinfo {year}
  {1994})}\BibitemShut {NoStop}%
\bibitem [{\citenamefont {Wang}\ \emph {et~al.}(2018)\citenamefont {Wang},
  \citenamefont {Xie}, \citenamefont {Suehiro}, \citenamefont {Takeda},\ and\
  \citenamefont {Hirosaki}}]{rjxie}%
  \BibitemOpen
  \bibfield  {author} {\bibinfo {author} {\bibfnamefont {L.}~\bibnamefont
  {Wang}}, \bibinfo {author} {\bibfnamefont {R.-J.}\ \bibnamefont {Xie}},
  \bibinfo {author} {\bibfnamefont {T.}~\bibnamefont {Suehiro}}, \bibinfo
  {author} {\bibfnamefont {T.}~\bibnamefont {Takeda}},\ and\ \bibinfo {author}
  {\bibfnamefont {N.}~\bibnamefont {Hirosaki}},\ }\bibfield  {title} {\bibinfo
  {title} {Down-conversion nitride materials for solid state lighting: recent
  advances and perspectives},\ }\href@noop {} {\bibfield  {journal} {\bibinfo
  {journal} {Chemical reviews}\ }\textbf {\bibinfo {volume} {118}},\ \bibinfo
  {pages} {1951} (\bibinfo {year} {2018})}\BibitemShut {NoStop}%
\bibitem [{\citenamefont {George}\ \emph {et~al.}(2013)\citenamefont {George},
  \citenamefont {Denault},\ and\ \citenamefont {Seshadri}}]{annurev}%
  \BibitemOpen
  \bibfield  {author} {\bibinfo {author} {\bibfnamefont {N.~C.}\ \bibnamefont
  {George}}, \bibinfo {author} {\bibfnamefont {K.~A.}\ \bibnamefont
  {Denault}},\ and\ \bibinfo {author} {\bibfnamefont {R.}~\bibnamefont
  {Seshadri}},\ }\bibfield  {title} {\bibinfo {title} {Phosphors for
  solid-state white lighting},\ }\href@noop {} {\bibfield  {journal} {\bibinfo
  {journal} {Annual Review of Materials Research}\ }\textbf {\bibinfo {volume}
  {43}},\ \bibinfo {pages} {481} (\bibinfo {year} {2013})}\BibitemShut
  {NoStop}%
\bibitem [{\citenamefont {Qin}\ \emph {et~al.}(2017)\citenamefont {Qin},
  \citenamefont {Liu}, \citenamefont {Huang}, \citenamefont {Bettinelli},\ and\
  \citenamefont {Liu}}]{lanthanide-review}%
  \BibitemOpen
  \bibfield  {author} {\bibinfo {author} {\bibfnamefont {X.}~\bibnamefont
  {Qin}}, \bibinfo {author} {\bibfnamefont {X.}~\bibnamefont {Liu}}, \bibinfo
  {author} {\bibfnamefont {W.}~\bibnamefont {Huang}}, \bibinfo {author}
  {\bibfnamefont {M.}~\bibnamefont {Bettinelli}},\ and\ \bibinfo {author}
  {\bibfnamefont {X.}~\bibnamefont {Liu}},\ }\bibfield  {title} {\bibinfo
  {title} {Lanthanide-activated phosphors based on 4f-5d optical transitions:
  theoretical and experimental aspects},\ }\href@noop {} {\bibfield  {journal}
  {\bibinfo  {journal} {Chemical reviews}\ }\textbf {\bibinfo {volume} {117}},\
  \bibinfo {pages} {4488} (\bibinfo {year} {2017})}\BibitemShut {NoStop}%
\bibitem [{\citenamefont {Lai}\ \emph {et~al.}(2022)\citenamefont {Lai},
  \citenamefont {Zhao}, \citenamefont {Zhao}, \citenamefont {Molokeev},\ and\
  \citenamefont {Xia}}]{eu-common}%
  \BibitemOpen
  \bibfield  {author} {\bibinfo {author} {\bibfnamefont {S.}~\bibnamefont
  {Lai}}, \bibinfo {author} {\bibfnamefont {M.}~\bibnamefont {Zhao}}, \bibinfo
  {author} {\bibfnamefont {Y.}~\bibnamefont {Zhao}}, \bibinfo {author}
  {\bibfnamefont {M.~S.}\ \bibnamefont {Molokeev}},\ and\ \bibinfo {author}
  {\bibfnamefont {Z.}~\bibnamefont {Xia}},\ }\bibfield  {title} {\bibinfo
  {title} {Eu2+ doping concentration-induced site-selective occupation and
  photoluminescence tuning in ksrscsi2o7: Eu2+ phosphor},\ }\href@noop {}
  {\bibfield  {journal} {\bibinfo  {journal} {ACS Materials Au}\ }\textbf
  {\bibinfo {volume} {2}},\ \bibinfo {pages} {374} (\bibinfo {year}
  {2022})}\BibitemShut {NoStop}%
\bibitem [{\citenamefont {Dorenbos}(2003)}]{300}%
  \BibitemOpen
  \bibfield  {author} {\bibinfo {author} {\bibfnamefont {P.}~\bibnamefont
  {Dorenbos}},\ }\bibfield  {title} {\bibinfo {title} {Energy of the first
  4f7→ 4f65d transition of eu2+ in inorganic compounds},\ }\href@noop {}
  {\bibfield  {journal} {\bibinfo  {journal} {Journal of luminescence}\
  }\textbf {\bibinfo {volume} {104}},\ \bibinfo {pages} {239} (\bibinfo {year}
  {2003})}\BibitemShut {NoStop}%
\bibitem [{\citenamefont {Lai}\ \emph {et~al.}(2020)\citenamefont {Lai},
  \citenamefont {Zhao}, \citenamefont {Qiao}, \citenamefont {Molokeev},\ and\
  \citenamefont {Xia}}]{data-driven}%
  \BibitemOpen
  \bibfield  {author} {\bibinfo {author} {\bibfnamefont {S.}~\bibnamefont
  {Lai}}, \bibinfo {author} {\bibfnamefont {M.}~\bibnamefont {Zhao}}, \bibinfo
  {author} {\bibfnamefont {J.}~\bibnamefont {Qiao}}, \bibinfo {author}
  {\bibfnamefont {M.~S.}\ \bibnamefont {Molokeev}},\ and\ \bibinfo {author}
  {\bibfnamefont {Z.}~\bibnamefont {Xia}},\ }\bibfield  {title} {\bibinfo
  {title} {Data-driven photoluminescence tuning in eu2+-doped phosphors},\
  }\href@noop {} {\bibfield  {journal} {\bibinfo  {journal} {The Journal of
  Physical Chemistry Letters}\ }\textbf {\bibinfo {volume} {11}},\ \bibinfo
  {pages} {5680} (\bibinfo {year} {2020})}\BibitemShut {NoStop}%
\bibitem [{\citenamefont {Jiang}\ \emph {et~al.}(2022)\citenamefont {Jiang},
  \citenamefont {Jiang}, \citenamefont {Lv},\ and\ \citenamefont
  {Su}}]{ml-phosphor}%
  \BibitemOpen
  \bibfield  {author} {\bibinfo {author} {\bibfnamefont {L.}~\bibnamefont
  {Jiang}}, \bibinfo {author} {\bibfnamefont {X.}~\bibnamefont {Jiang}},
  \bibinfo {author} {\bibfnamefont {G.}~\bibnamefont {Lv}},\ and\ \bibinfo
  {author} {\bibfnamefont {Y.}~\bibnamefont {Su}},\ }\bibfield  {title}
  {\bibinfo {title} {A mini review of machine learning in inorganic
  phosphors},\ }\href@noop {} {\bibfield  {journal} {\bibinfo  {journal}
  {Journal of Materials Informatics}\ }\textbf {\bibinfo {volume} {2}},\
  \bibinfo {pages} {14} (\bibinfo {year} {2022})}\BibitemShut {NoStop}%
\bibitem [{\citenamefont {Barai}\ and\ \citenamefont
  {Dhoble}(2019)}]{lasso-ann}%
  \BibitemOpen
  \bibfield  {author} {\bibinfo {author} {\bibfnamefont {V.~L.}\ \bibnamefont
  {Barai}}\ and\ \bibinfo {author} {\bibfnamefont {S.}~\bibnamefont {Dhoble}},\
  }\bibfield  {title} {\bibinfo {title} {Prediction of excitation wavelength of
  phosphors by using machine learning model},\ }\href@noop {} {\bibfield
  {journal} {\bibinfo  {journal} {Journal of Luminescence}\ }\textbf {\bibinfo
  {volume} {208}},\ \bibinfo {pages} {437} (\bibinfo {year}
  {2019})}\BibitemShut {NoStop}%
\bibitem [{\citenamefont {Koyama}\ \emph {et~al.}(2023)\citenamefont {Koyama},
  \citenamefont {Ikeno}, \citenamefont {Harada}, \citenamefont {Funahashi},
  \citenamefont {Takeda},\ and\ \citenamefont {Hirosaki}}]{too-close}%
  \BibitemOpen
  \bibfield  {author} {\bibinfo {author} {\bibfnamefont {Y.}~\bibnamefont
  {Koyama}}, \bibinfo {author} {\bibfnamefont {H.}~\bibnamefont {Ikeno}},
  \bibinfo {author} {\bibfnamefont {M.}~\bibnamefont {Harada}}, \bibinfo
  {author} {\bibfnamefont {S.}~\bibnamefont {Funahashi}}, \bibinfo {author}
  {\bibfnamefont {T.}~\bibnamefont {Takeda}},\ and\ \bibinfo {author}
  {\bibfnamefont {N.}~\bibnamefont {Hirosaki}},\ }\bibfield  {title} {\bibinfo
  {title} {Rapid discovery of new eu 2+-activated phosphors with a designed
  luminescence color using a data-driven approach},\ }\href@noop {} {\bibfield
  {journal} {\bibinfo  {journal} {Materials Advances}\ }\textbf {\bibinfo
  {volume} {4}},\ \bibinfo {pages} {231} (\bibinfo {year} {2023})}\BibitemShut
  {NoStop}%
\bibitem [{\citenamefont {Kim}\ \emph {et~al.}(2023)\citenamefont {Kim},
  \citenamefont {Jurakuziev}, \citenamefont {Akhtar},\ and\ \citenamefont
  {Yang}}]{rf}%
  \BibitemOpen
  \bibfield  {author} {\bibinfo {author} {\bibfnamefont {T.-G.}\ \bibnamefont
  {Kim}}, \bibinfo {author} {\bibfnamefont {D.}~\bibnamefont {Jurakuziev}},
  \bibinfo {author} {\bibfnamefont {M.~S.}\ \bibnamefont {Akhtar}},\ and\
  \bibinfo {author} {\bibfnamefont {O.-B.}\ \bibnamefont {Yang}},\ }\bibfield
  {title} {\bibinfo {title} {Machine learning investigation to predict the
  relationship between photoluminescence and crystalline properties of blue
  phosphor ba0. 9-xsrxmgal10o17: Eu2+},\ }\href@noop {} {\bibfield  {journal}
  {\bibinfo  {journal} {Journal of Science: Advanced Materials and Devices}\
  }\textbf {\bibinfo {volume} {8}},\ \bibinfo {pages} {100550} (\bibinfo {year}
  {2023})}\BibitemShut {NoStop}%
\bibitem [{\citenamefont {Gori}\ \emph {et~al.}(2005)\citenamefont {Gori},
  \citenamefont {Monfardini},\ and\ \citenamefont {Scarselli}}]{gnn}%
  \BibitemOpen
  \bibfield  {author} {\bibinfo {author} {\bibfnamefont {M.}~\bibnamefont
  {Gori}}, \bibinfo {author} {\bibfnamefont {G.}~\bibnamefont {Monfardini}},\
  and\ \bibinfo {author} {\bibfnamefont {F.}~\bibnamefont {Scarselli}},\
  }\bibfield  {title} {\bibinfo {title} {A new model for learning in graph
  domains},\ }in\ \href@noop {} {\emph {\bibinfo {booktitle} {Proceedings. 2005
  IEEE International Joint Conference on Neural Networks, 2005.}}},\
  Vol.~\bibinfo {volume} {2}\ (\bibinfo {organization} {IEEE},\ \bibinfo {year}
  {2005})\ pp.\ \bibinfo {pages} {729--734}\BibitemShut {NoStop}%
\bibitem [{\citenamefont {Xie}\ and\ \citenamefont {Grossman}(2018)}]{cgcnn}%
  \BibitemOpen
  \bibfield  {author} {\bibinfo {author} {\bibfnamefont {T.}~\bibnamefont
  {Xie}}\ and\ \bibinfo {author} {\bibfnamefont {J.~C.}\ \bibnamefont
  {Grossman}},\ }\bibfield  {title} {\bibinfo {title} {Crystal graph
  convolutional neural networks for an accurate and interpretable prediction of
  material properties},\ }\href@noop {} {\bibfield  {journal} {\bibinfo
  {journal} {Physical review letters}\ }\textbf {\bibinfo {volume} {120}},\
  \bibinfo {pages} {145301} (\bibinfo {year} {2018})}\BibitemShut {NoStop}%
\bibitem [{\citenamefont {Reiser}\ \emph {et~al.}(2022)\citenamefont {Reiser},
  \citenamefont {Neubert}, \citenamefont {Eberhard}, \citenamefont {Torresi},
  \citenamefont {Zhou}, \citenamefont {Shao}, \citenamefont {Metni},
  \citenamefont {van Hoesel}, \citenamefont {Schopmans}, \citenamefont {Sommer}
  \emph {et~al.}}]{gnn-review}%
  \BibitemOpen
  \bibfield  {author} {\bibinfo {author} {\bibfnamefont {P.}~\bibnamefont
  {Reiser}}, \bibinfo {author} {\bibfnamefont {M.}~\bibnamefont {Neubert}},
  \bibinfo {author} {\bibfnamefont {A.}~\bibnamefont {Eberhard}}, \bibinfo
  {author} {\bibfnamefont {L.}~\bibnamefont {Torresi}}, \bibinfo {author}
  {\bibfnamefont {C.}~\bibnamefont {Zhou}}, \bibinfo {author} {\bibfnamefont
  {C.}~\bibnamefont {Shao}}, \bibinfo {author} {\bibfnamefont {H.}~\bibnamefont
  {Metni}}, \bibinfo {author} {\bibfnamefont {C.}~\bibnamefont {van Hoesel}},
  \bibinfo {author} {\bibfnamefont {H.}~\bibnamefont {Schopmans}}, \bibinfo
  {author} {\bibfnamefont {T.}~\bibnamefont {Sommer}}, \emph {et~al.},\
  }\bibfield  {title} {\bibinfo {title} {Graph neural networks for materials
  science and chemistry},\ }\href@noop {} {\bibfield  {journal} {\bibinfo
  {journal} {Communications Materials}\ }\textbf {\bibinfo {volume} {3}},\
  \bibinfo {pages} {93} (\bibinfo {year} {2022})}\BibitemShut {NoStop}%
\bibitem [{\citenamefont {Zagorac}\ \emph {et~al.}(2019)\citenamefont
  {Zagorac}, \citenamefont {M{\"u}ller}, \citenamefont {Ruehl}, \citenamefont
  {Zagorac},\ and\ \citenamefont {Rehme}}]{icsd}%
  \BibitemOpen
  \bibfield  {author} {\bibinfo {author} {\bibfnamefont {D.}~\bibnamefont
  {Zagorac}}, \bibinfo {author} {\bibfnamefont {H.}~\bibnamefont {M{\"u}ller}},
  \bibinfo {author} {\bibfnamefont {S.}~\bibnamefont {Ruehl}}, \bibinfo
  {author} {\bibfnamefont {J.}~\bibnamefont {Zagorac}},\ and\ \bibinfo {author}
  {\bibfnamefont {S.}~\bibnamefont {Rehme}},\ }\bibfield  {title} {\bibinfo
  {title} {Recent developments in the inorganic crystal structure database:
  theoretical crystal structure data and related features},\ }\href@noop {}
  {\bibfield  {journal} {\bibinfo  {journal} {Journal of applied
  crystallography}\ }\textbf {\bibinfo {volume} {52}},\ \bibinfo {pages} {918}
  (\bibinfo {year} {2019})}\BibitemShut {NoStop}%
\bibitem [{\citenamefont {Jain}\ \emph {et~al.}(2013)\citenamefont {Jain},
  \citenamefont {Ong}, \citenamefont {Hautier}, \citenamefont {Chen},
  \citenamefont {Richards}, \citenamefont {Dacek}, \citenamefont {Cholia},
  \citenamefont {Gunter}, \citenamefont {Skinner}, \citenamefont {Ceder} \emph
  {et~al.}}]{matproj}%
  \BibitemOpen
  \bibfield  {author} {\bibinfo {author} {\bibfnamefont {A.}~\bibnamefont
  {Jain}}, \bibinfo {author} {\bibfnamefont {S.~P.}\ \bibnamefont {Ong}},
  \bibinfo {author} {\bibfnamefont {G.}~\bibnamefont {Hautier}}, \bibinfo
  {author} {\bibfnamefont {W.}~\bibnamefont {Chen}}, \bibinfo {author}
  {\bibfnamefont {W.~D.}\ \bibnamefont {Richards}}, \bibinfo {author}
  {\bibfnamefont {S.}~\bibnamefont {Dacek}}, \bibinfo {author} {\bibfnamefont
  {S.}~\bibnamefont {Cholia}}, \bibinfo {author} {\bibfnamefont
  {D.}~\bibnamefont {Gunter}}, \bibinfo {author} {\bibfnamefont
  {D.}~\bibnamefont {Skinner}}, \bibinfo {author} {\bibfnamefont
  {G.}~\bibnamefont {Ceder}}, \emph {et~al.},\ }\bibfield  {title} {\bibinfo
  {title} {Commentary: The materials project: A materials genome approach to
  accelerating materials innovation},\ }\href@noop {} {\bibfield  {journal}
  {\bibinfo  {journal} {APL materials}\ }\textbf {\bibinfo {volume} {1}},\
  \bibinfo {pages} {011002} (\bibinfo {year} {2013})}\BibitemShut {NoStop}%
\bibitem [{\citenamefont {Saal}\ \emph {et~al.}(2013)\citenamefont {Saal},
  \citenamefont {Kirklin}, \citenamefont {Aykol}, \citenamefont {Meredig},\
  and\ \citenamefont {Wolverton}}]{oqmd}%
  \BibitemOpen
  \bibfield  {author} {\bibinfo {author} {\bibfnamefont {J.~E.}\ \bibnamefont
  {Saal}}, \bibinfo {author} {\bibfnamefont {S.}~\bibnamefont {Kirklin}},
  \bibinfo {author} {\bibfnamefont {M.}~\bibnamefont {Aykol}}, \bibinfo
  {author} {\bibfnamefont {B.}~\bibnamefont {Meredig}},\ and\ \bibinfo {author}
  {\bibfnamefont {C.}~\bibnamefont {Wolverton}},\ }\bibfield  {title} {\bibinfo
  {title} {Materials design and discovery with high-throughput density
  functional theory: the open quantum materials database (oqmd)},\ }\href@noop
  {} {\bibfield  {journal} {\bibinfo  {journal} {Jom}\ }\textbf {\bibinfo
  {volume} {65}},\ \bibinfo {pages} {1501} (\bibinfo {year}
  {2013})}\BibitemShut {NoStop}%
\bibitem [{\citenamefont {Bruna}\ \emph {et~al.}(2014)\citenamefont {Bruna},
  \citenamefont {Zaremba}, \citenamefont {Szlam},\ and\ \citenamefont
  {Lecun}}]{spectral-conv}%
  \BibitemOpen
  \bibfield  {author} {\bibinfo {author} {\bibfnamefont {J.}~\bibnamefont
  {Bruna}}, \bibinfo {author} {\bibfnamefont {W.}~\bibnamefont {Zaremba}},
  \bibinfo {author} {\bibfnamefont {A.}~\bibnamefont {Szlam}},\ and\ \bibinfo
  {author} {\bibfnamefont {Y.}~\bibnamefont {Lecun}},\ }\bibfield  {title}
  {\bibinfo {title} {Spectral networks and locally connected networks on
  graphs},\ }in\ \href@noop {} {\emph {\bibinfo {booktitle} {International
  Conference on Learning Representations (ICLR2014), CBLS, April 2014}}}\
  (\bibinfo {year} {2014})\BibitemShut {NoStop}%
\bibitem [{\citenamefont {Kipf}\ and\ \citenamefont {Welling}(2017)}]{gcnn}%
  \BibitemOpen
  \bibfield  {author} {\bibinfo {author} {\bibfnamefont {T.~N.}\ \bibnamefont
  {Kipf}}\ and\ \bibinfo {author} {\bibfnamefont {M.}~\bibnamefont {Welling}},\
  }\bibfield  {title} {\bibinfo {title} {Semi-supervised classification with
  graph convolutional networks},\ }in\ \href
  {https://openreview.net/forum?id=SJU4ayYgl} {\emph {\bibinfo {booktitle}
  {International Conference on Learning Representations}}}\ (\bibinfo {year}
  {2017})\BibitemShut {NoStop}%
\bibitem [{\citenamefont {Gilmer}\ \emph {et~al.}(2017)\citenamefont {Gilmer},
  \citenamefont {Schoenholz}, \citenamefont {Riley}, \citenamefont {Vinyals},\
  and\ \citenamefont {Dahl}}]{mpnn}%
  \BibitemOpen
  \bibfield  {author} {\bibinfo {author} {\bibfnamefont {J.}~\bibnamefont
  {Gilmer}}, \bibinfo {author} {\bibfnamefont {S.~S.}\ \bibnamefont
  {Schoenholz}}, \bibinfo {author} {\bibfnamefont {P.~F.}\ \bibnamefont
  {Riley}}, \bibinfo {author} {\bibfnamefont {O.}~\bibnamefont {Vinyals}},\
  and\ \bibinfo {author} {\bibfnamefont {G.~E.}\ \bibnamefont {Dahl}},\
  }\bibfield  {title} {\bibinfo {title} {Neural message passing for quantum
  chemistry},\ }in\ \href@noop {} {\emph {\bibinfo {booktitle} {International
  conference on machine learning}}}\ (\bibinfo {organization} {PMLR},\ \bibinfo
  {year} {2017})\ pp.\ \bibinfo {pages} {1263--1272}\BibitemShut {NoStop}%
\bibitem [{\citenamefont {Fung}\ \emph {et~al.}(2021)\citenamefont {Fung},
  \citenamefont {Zhang}, \citenamefont {Juarez},\ and\ \citenamefont
  {Sumpter}}]{gnn-benchmark}%
  \BibitemOpen
  \bibfield  {author} {\bibinfo {author} {\bibfnamefont {V.}~\bibnamefont
  {Fung}}, \bibinfo {author} {\bibfnamefont {J.}~\bibnamefont {Zhang}},
  \bibinfo {author} {\bibfnamefont {E.}~\bibnamefont {Juarez}},\ and\ \bibinfo
  {author} {\bibfnamefont {B.~G.}\ \bibnamefont {Sumpter}},\ }\bibfield
  {title} {\bibinfo {title} {Benchmarking graph neural networks for materials
  chemistry},\ }\href@noop {} {\bibfield  {journal} {\bibinfo  {journal} {npj
  Computational Materials}\ }\textbf {\bibinfo {volume} {7}},\ \bibinfo {pages}
  {84} (\bibinfo {year} {2021})}\BibitemShut {NoStop}%
\bibitem [{\citenamefont {Krallinger}\ \emph {et~al.}(2017)\citenamefont
  {Krallinger}, \citenamefont {Rabal}, \citenamefont {Lourenco}, \citenamefont
  {Oyarzabal},\ and\ \citenamefont {Valencia}}]{ceder-review-1}%
  \BibitemOpen
  \bibfield  {author} {\bibinfo {author} {\bibfnamefont {M.}~\bibnamefont
  {Krallinger}}, \bibinfo {author} {\bibfnamefont {O.}~\bibnamefont {Rabal}},
  \bibinfo {author} {\bibfnamefont {A.}~\bibnamefont {Lourenco}}, \bibinfo
  {author} {\bibfnamefont {J.}~\bibnamefont {Oyarzabal}},\ and\ \bibinfo
  {author} {\bibfnamefont {A.}~\bibnamefont {Valencia}},\ }\bibfield  {title}
  {\bibinfo {title} {Information retrieval and text mining technologies for
  chemistry},\ }\href@noop {} {\bibfield  {journal} {\bibinfo  {journal}
  {Chemical reviews}\ }\textbf {\bibinfo {volume} {117}},\ \bibinfo {pages}
  {7673} (\bibinfo {year} {2017})}\BibitemShut {NoStop}%
\bibitem [{\citenamefont {Kononova}\ \emph {et~al.}(2021)\citenamefont
  {Kononova}, \citenamefont {He}, \citenamefont {Huo}, \citenamefont
  {Trewartha}, \citenamefont {Olivetti},\ and\ \citenamefont
  {Ceder}}]{ceder-review-2}%
  \BibitemOpen
  \bibfield  {author} {\bibinfo {author} {\bibfnamefont {O.}~\bibnamefont
  {Kononova}}, \bibinfo {author} {\bibfnamefont {T.}~\bibnamefont {He}},
  \bibinfo {author} {\bibfnamefont {H.}~\bibnamefont {Huo}}, \bibinfo {author}
  {\bibfnamefont {A.}~\bibnamefont {Trewartha}}, \bibinfo {author}
  {\bibfnamefont {E.~A.}\ \bibnamefont {Olivetti}},\ and\ \bibinfo {author}
  {\bibfnamefont {G.}~\bibnamefont {Ceder}},\ }\bibfield  {title} {\bibinfo
  {title} {Opportunities and challenges of text mining in materials research},\
  }\href@noop {} {\bibfield  {journal} {\bibinfo  {journal} {Iscience}\
  }\textbf {\bibinfo {volume} {24}},\ \bibinfo {pages} {102155} (\bibinfo
  {year} {2021})}\BibitemShut {NoStop}%
\bibitem [{\citenamefont {Kononova}\ \emph {et~al.}(2019)\citenamefont
  {Kononova}, \citenamefont {Huo}, \citenamefont {He}, \citenamefont {Rong},
  \citenamefont {Botari}, \citenamefont {Sun}, \citenamefont {Tshitoyan},\ and\
  \citenamefont {Ceder}}]{ceder}%
  \BibitemOpen
  \bibfield  {author} {\bibinfo {author} {\bibfnamefont {O.}~\bibnamefont
  {Kononova}}, \bibinfo {author} {\bibfnamefont {H.}~\bibnamefont {Huo}},
  \bibinfo {author} {\bibfnamefont {T.}~\bibnamefont {He}}, \bibinfo {author}
  {\bibfnamefont {Z.}~\bibnamefont {Rong}}, \bibinfo {author} {\bibfnamefont
  {T.}~\bibnamefont {Botari}}, \bibinfo {author} {\bibfnamefont
  {W.}~\bibnamefont {Sun}}, \bibinfo {author} {\bibfnamefont {V.}~\bibnamefont
  {Tshitoyan}},\ and\ \bibinfo {author} {\bibfnamefont {G.}~\bibnamefont
  {Ceder}},\ }\bibfield  {title} {\bibinfo {title} {Text-mined dataset of
  inorganic materials synthesis recipes},\ }\href@noop {} {\bibfield  {journal}
  {\bibinfo  {journal} {Scientific data}\ }\textbf {\bibinfo {volume} {6}},\
  \bibinfo {pages} {203} (\bibinfo {year} {2019})}\BibitemShut {NoStop}%
\bibitem [{\citenamefont {Jensen}\ \emph {et~al.}(2021)\citenamefont {Jensen},
  \citenamefont {Kwon}, \citenamefont {Schwalbe-Koda}, \citenamefont {Paris},
  \citenamefont {G{\'o}mez-Bombarelli}, \citenamefont {Rom{\'a}n-Leshkov},
  \citenamefont {Corma}, \citenamefont {Moliner},\ and\ \citenamefont
  {Olivetti}}]{zeolite}%
  \BibitemOpen
  \bibfield  {author} {\bibinfo {author} {\bibfnamefont {Z.}~\bibnamefont
  {Jensen}}, \bibinfo {author} {\bibfnamefont {S.}~\bibnamefont {Kwon}},
  \bibinfo {author} {\bibfnamefont {D.}~\bibnamefont {Schwalbe-Koda}}, \bibinfo
  {author} {\bibfnamefont {C.}~\bibnamefont {Paris}}, \bibinfo {author}
  {\bibfnamefont {R.}~\bibnamefont {G{\'o}mez-Bombarelli}}, \bibinfo {author}
  {\bibfnamefont {Y.}~\bibnamefont {Rom{\'a}n-Leshkov}}, \bibinfo {author}
  {\bibfnamefont {A.}~\bibnamefont {Corma}}, \bibinfo {author} {\bibfnamefont
  {M.}~\bibnamefont {Moliner}},\ and\ \bibinfo {author} {\bibfnamefont {E.~A.}\
  \bibnamefont {Olivetti}},\ }\bibfield  {title} {\bibinfo {title} {Discovering
  relationships between osdas and zeolites through data mining and generative
  neural networks},\ }\href@noop {} {\bibfield  {journal} {\bibinfo  {journal}
  {ACS Central Science}\ }\textbf {\bibinfo {volume} {7}},\ \bibinfo {pages}
  {858} (\bibinfo {year} {2021})}\BibitemShut {NoStop}%
\bibitem [{\citenamefont {Swain}\ and\ \citenamefont
  {Cole}(2016)}]{chemdataextractor}%
  \BibitemOpen
  \bibfield  {author} {\bibinfo {author} {\bibfnamefont {M.~C.}\ \bibnamefont
  {Swain}}\ and\ \bibinfo {author} {\bibfnamefont {J.~M.}\ \bibnamefont
  {Cole}},\ }\bibfield  {title} {\bibinfo {title} {Chemdataextractor: a toolkit
  for automated extraction of chemical information from the scientific
  literature},\ }\href@noop {} {\bibfield  {journal} {\bibinfo  {journal}
  {Journal of chemical information and modeling}\ }\textbf {\bibinfo {volume}
  {56}},\ \bibinfo {pages} {1894} (\bibinfo {year} {2016})}\BibitemShut
  {NoStop}%
\bibitem [{\citenamefont {Jessop}\ \emph {et~al.}(2011)\citenamefont {Jessop},
  \citenamefont {Adams}, \citenamefont {Willighagen}, \citenamefont {Hawizy},\
  and\ \citenamefont {Murray-Rust}}]{oscar4}%
  \BibitemOpen
  \bibfield  {author} {\bibinfo {author} {\bibfnamefont {D.~M.}\ \bibnamefont
  {Jessop}}, \bibinfo {author} {\bibfnamefont {S.~E.}\ \bibnamefont {Adams}},
  \bibinfo {author} {\bibfnamefont {E.~L.}\ \bibnamefont {Willighagen}},
  \bibinfo {author} {\bibfnamefont {L.}~\bibnamefont {Hawizy}},\ and\ \bibinfo
  {author} {\bibfnamefont {P.}~\bibnamefont {Murray-Rust}},\ }\bibfield
  {title} {\bibinfo {title} {Oscar4: a flexible architecture for chemical
  text-mining},\ }\href@noop {} {\bibfield  {journal} {\bibinfo  {journal}
  {Journal of cheminformatics}\ }\textbf {\bibinfo {volume} {3}},\ \bibinfo
  {pages} {1} (\bibinfo {year} {2011})}\BibitemShut {NoStop}%
\bibitem [{\citenamefont {Hawizy}\ \emph {et~al.}(2011)\citenamefont {Hawizy},
  \citenamefont {Jessop}, \citenamefont {Adams},\ and\ \citenamefont
  {Murray-Rust}}]{chemtagger}%
  \BibitemOpen
  \bibfield  {author} {\bibinfo {author} {\bibfnamefont {L.}~\bibnamefont
  {Hawizy}}, \bibinfo {author} {\bibfnamefont {D.~M.}\ \bibnamefont {Jessop}},
  \bibinfo {author} {\bibfnamefont {N.}~\bibnamefont {Adams}},\ and\ \bibinfo
  {author} {\bibfnamefont {P.}~\bibnamefont {Murray-Rust}},\ }\bibfield
  {title} {\bibinfo {title} {Chemicaltagger: A tool for semantic text-mining in
  chemistry},\ }\href@noop {} {\bibfield  {journal} {\bibinfo  {journal}
  {Journal of cheminformatics}\ }\textbf {\bibinfo {volume} {3}},\ \bibinfo
  {pages} {1} (\bibinfo {year} {2011})}\BibitemShut {NoStop}%
\bibitem [{\citenamefont {Manning}(2022)}]{nlp-to-llm}%
  \BibitemOpen
  \bibfield  {author} {\bibinfo {author} {\bibfnamefont {C.~D.}\ \bibnamefont
  {Manning}},\ }\bibfield  {title} {\bibinfo {title} {Human language
  understanding \& reasoning},\ }\href@noop {} {\bibfield  {journal} {\bibinfo
  {journal} {Daedalus}\ }\textbf {\bibinfo {volume} {151}},\ \bibinfo {pages}
  {127} (\bibinfo {year} {2022})}\BibitemShut {NoStop}%
\bibitem [{\citenamefont {Polak}\ and\ \citenamefont
  {Morgan}(2023)}]{gpt-extract-1}%
  \BibitemOpen
  \bibfield  {author} {\bibinfo {author} {\bibfnamefont {M.~P.}\ \bibnamefont
  {Polak}}\ and\ \bibinfo {author} {\bibfnamefont {D.}~\bibnamefont {Morgan}},\
  }\bibfield  {title} {\bibinfo {title} {Extracting accurate materials data
  from research papers with conversational language models and prompt
  engineering--example of chatgpt},\ }\href@noop {} {\bibfield  {journal}
  {\bibinfo  {journal} {arXiv preprint arXiv:2303.05352}\ } (\bibinfo {year}
  {2023})}\BibitemShut {NoStop}%
\bibitem [{\citenamefont {Xiao}\ \emph {et~al.}(2023)\citenamefont {Xiao},
  \citenamefont {Li}, \citenamefont {Moon}, \citenamefont {Roell},
  \citenamefont {Chen},\ and\ \citenamefont {Tang}}]{bioxiv}%
  \BibitemOpen
  \bibfield  {author} {\bibinfo {author} {\bibfnamefont {Z.}~\bibnamefont
  {Xiao}}, \bibinfo {author} {\bibfnamefont {W.}~\bibnamefont {Li}}, \bibinfo
  {author} {\bibfnamefont {H.}~\bibnamefont {Moon}}, \bibinfo {author}
  {\bibfnamefont {G.~W.}\ \bibnamefont {Roell}}, \bibinfo {author}
  {\bibfnamefont {Y.}~\bibnamefont {Chen}},\ and\ \bibinfo {author}
  {\bibfnamefont {Y.~J.}\ \bibnamefont {Tang}},\ }\bibfield  {title} {\bibinfo
  {title} {Generative artificial intelligence gpt-4 accelerates knowledge
  mining and machine learning for synthetic biology},\ }\href@noop {}
  {\bibfield  {journal} {\bibinfo  {journal} {bioRxiv}\ ,\ \bibinfo {pages}
  {2023}} (\bibinfo {year} {2023})}\BibitemShut {NoStop}%
\bibitem [{\citenamefont {Radford}\ \emph {et~al.}(2018)\citenamefont
  {Radford}, \citenamefont {Narasimhan}, \citenamefont {Salimans},
  \citenamefont {Sutskever} \emph {et~al.}}]{gpt1}%
  \BibitemOpen
  \bibfield  {author} {\bibinfo {author} {\bibfnamefont {A.}~\bibnamefont
  {Radford}}, \bibinfo {author} {\bibfnamefont {K.}~\bibnamefont {Narasimhan}},
  \bibinfo {author} {\bibfnamefont {T.}~\bibnamefont {Salimans}}, \bibinfo
  {author} {\bibfnamefont {I.}~\bibnamefont {Sutskever}}, \emph {et~al.},\
  }\bibfield  {title} {\bibinfo {title} {Improving language understanding by
  generative pre-training},\ }\href@noop {} {\  (\bibinfo {year}
  {2018})}\BibitemShut {NoStop}%
\bibitem [{\citenamefont {Kenton}\ and\ \citenamefont
  {Toutanova}(2019)}]{bert}%
  \BibitemOpen
  \bibfield  {author} {\bibinfo {author} {\bibfnamefont {J.~D. M.-W.~C.}\
  \bibnamefont {Kenton}}\ and\ \bibinfo {author} {\bibfnamefont {L.~K.}\
  \bibnamefont {Toutanova}},\ }\bibfield  {title} {\bibinfo {title} {Bert:
  Pre-training of deep bidirectional transformers for language understanding},\
  }in\ \href@noop {} {\emph {\bibinfo {booktitle} {Proceedings of
  naacL-HLT}}},\ Vol.~\bibinfo {volume} {1}\ (\bibinfo {year} {2019})\
  p.~\bibinfo {pages} {2}\BibitemShut {NoStop}%
\bibitem [{\citenamefont {Brown}\ \emph {et~al.}(2020)\citenamefont {Brown},
  \citenamefont {Mann}, \citenamefont {Ryder}, \citenamefont {Subbiah},
  \citenamefont {Kaplan}, \citenamefont {Dhariwal}, \citenamefont
  {Neelakantan}, \citenamefont {Shyam}, \citenamefont {Sastry}, \citenamefont
  {Askell} \emph {et~al.}}]{gpt3}%
  \BibitemOpen
  \bibfield  {author} {\bibinfo {author} {\bibfnamefont {T.}~\bibnamefont
  {Brown}}, \bibinfo {author} {\bibfnamefont {B.}~\bibnamefont {Mann}},
  \bibinfo {author} {\bibfnamefont {N.}~\bibnamefont {Ryder}}, \bibinfo
  {author} {\bibfnamefont {M.}~\bibnamefont {Subbiah}}, \bibinfo {author}
  {\bibfnamefont {J.~D.}\ \bibnamefont {Kaplan}}, \bibinfo {author}
  {\bibfnamefont {P.}~\bibnamefont {Dhariwal}}, \bibinfo {author}
  {\bibfnamefont {A.}~\bibnamefont {Neelakantan}}, \bibinfo {author}
  {\bibfnamefont {P.}~\bibnamefont {Shyam}}, \bibinfo {author} {\bibfnamefont
  {G.}~\bibnamefont {Sastry}}, \bibinfo {author} {\bibfnamefont
  {A.}~\bibnamefont {Askell}}, \emph {et~al.},\ }\bibfield  {title} {\bibinfo
  {title} {Language models are few-shot learners},\ }\href@noop {} {\bibfield
  {journal} {\bibinfo  {journal} {Advances in neural information processing
  systems}\ }\textbf {\bibinfo {volume} {33}},\ \bibinfo {pages} {1877}
  (\bibinfo {year} {2020})}\BibitemShut {NoStop}%
\bibitem [{\citenamefont {Bahdanau}\ \emph {et~al.}(2014)\citenamefont
  {Bahdanau}, \citenamefont {Cho},\ and\ \citenamefont {Bengio}}]{attention}%
  \BibitemOpen
  \bibfield  {author} {\bibinfo {author} {\bibfnamefont {D.}~\bibnamefont
  {Bahdanau}}, \bibinfo {author} {\bibfnamefont {K.}~\bibnamefont {Cho}},\ and\
  \bibinfo {author} {\bibfnamefont {Y.}~\bibnamefont {Bengio}},\ }\bibfield
  {title} {\bibinfo {title} {Neural machine translation by jointly learning to
  align and translate},\ }\href@noop {} {\bibfield  {journal} {\bibinfo
  {journal} {arXiv preprint arXiv:1409.0473}\ } (\bibinfo {year}
  {2014})}\BibitemShut {NoStop}%
\bibitem [{\citenamefont {Vaswani}\ \emph {et~al.}(2017)\citenamefont
  {Vaswani}, \citenamefont {Shazeer}, \citenamefont {Parmar}, \citenamefont
  {Uszkoreit}, \citenamefont {Jones}, \citenamefont {Gomez}, \citenamefont
  {Kaiser},\ and\ \citenamefont {Polosukhin}}]{transformer}%
  \BibitemOpen
  \bibfield  {author} {\bibinfo {author} {\bibfnamefont {A.}~\bibnamefont
  {Vaswani}}, \bibinfo {author} {\bibfnamefont {N.}~\bibnamefont {Shazeer}},
  \bibinfo {author} {\bibfnamefont {N.}~\bibnamefont {Parmar}}, \bibinfo
  {author} {\bibfnamefont {J.}~\bibnamefont {Uszkoreit}}, \bibinfo {author}
  {\bibfnamefont {L.}~\bibnamefont {Jones}}, \bibinfo {author} {\bibfnamefont
  {A.~N.}\ \bibnamefont {Gomez}}, \bibinfo {author} {\bibfnamefont
  {{\L}.}~\bibnamefont {Kaiser}},\ and\ \bibinfo {author} {\bibfnamefont
  {I.}~\bibnamefont {Polosukhin}},\ }\bibfield  {title} {\bibinfo {title}
  {Attention is all you need},\ }\href@noop {} {\bibfield  {journal} {\bibinfo
  {journal} {Advances in neural information processing systems}\ }\textbf
  {\bibinfo {volume} {30}} (\bibinfo {year} {2017})}\BibitemShut {NoStop}%
\bibitem [{\citenamefont {Hochreiter}\ and\ \citenamefont
  {Schmidhuber}(1997)}]{lstm1997}%
  \BibitemOpen
  \bibfield  {author} {\bibinfo {author} {\bibfnamefont {S.}~\bibnamefont
  {Hochreiter}}\ and\ \bibinfo {author} {\bibfnamefont {J.}~\bibnamefont
  {Schmidhuber}},\ }\bibfield  {title} {\bibinfo {title} {Long short-term
  memory},\ }\href@noop {} {\bibfield  {journal} {\bibinfo  {journal} {Neural
  computation}\ }\textbf {\bibinfo {volume} {9}},\ \bibinfo {pages} {1735}
  (\bibinfo {year} {1997})}\BibitemShut {NoStop}%
\bibitem [{\citenamefont {Radford}\ \emph {et~al.}(2019)\citenamefont
  {Radford}, \citenamefont {Wu}, \citenamefont {Child}, \citenamefont {Luan},
  \citenamefont {Amodei}, \citenamefont {Sutskever} \emph {et~al.}}]{gpt2}%
  \BibitemOpen
  \bibfield  {author} {\bibinfo {author} {\bibfnamefont {A.}~\bibnamefont
  {Radford}}, \bibinfo {author} {\bibfnamefont {J.}~\bibnamefont {Wu}},
  \bibinfo {author} {\bibfnamefont {R.}~\bibnamefont {Child}}, \bibinfo
  {author} {\bibfnamefont {D.}~\bibnamefont {Luan}}, \bibinfo {author}
  {\bibfnamefont {D.}~\bibnamefont {Amodei}}, \bibinfo {author} {\bibfnamefont
  {I.}~\bibnamefont {Sutskever}}, \emph {et~al.},\ }\bibfield  {title}
  {\bibinfo {title} {Language models are unsupervised multitask learners},\
  }\href@noop {} {\bibfield  {journal} {\bibinfo  {journal} {OpenAI blog}\
  }\textbf {\bibinfo {volume} {1}},\ \bibinfo {pages} {9} (\bibinfo {year}
  {2019})}\BibitemShut {NoStop}%
\bibitem [{\citenamefont {OpenAI}(2023)}]{gpt4}%
  \BibitemOpen
  \bibfield  {author} {\bibinfo {author} {\bibnamefont {OpenAI}},\ }\href@noop
  {} {\bibinfo {title} {Gpt-4 technical report}} (\bibinfo {year} {2023}),\
  \Eprint {https://arxiv.org/abs/2303.08774} {arXiv:2303.08774 [cs.CL]}
  \BibitemShut {NoStop}%
\bibitem [{\citenamefont {Knox}\ and\ \citenamefont {Stone}(2011)}]{rlhf-icml}%
  \BibitemOpen
  \bibfield  {author} {\bibinfo {author} {\bibfnamefont {W.~B.}\ \bibnamefont
  {Knox}}\ and\ \bibinfo {author} {\bibfnamefont {P.}~\bibnamefont {Stone}},\
  }\bibfield  {title} {\bibinfo {title} {Augmenting reinforcement learning with
  human feedback},\ }in\ \href@noop {} {\emph {\bibinfo {booktitle} {ICML 2011
  Workshop on New Developments in Imitation Learning (July 2011)}}},\ Vol.\
  \bibinfo {volume} {855}\ (\bibinfo {year} {2011})\ p.~\bibinfo {pages}
  {3}\BibitemShut {NoStop}%
\bibitem [{\citenamefont {Bennie}\ and\ \citenamefont
  {Erickson}(2023)}]{messy}%
  \BibitemOpen
  \bibfield  {author} {\bibinfo {author} {\bibfnamefont {B.}~\bibnamefont
  {Bennie}}\ and\ \bibinfo {author} {\bibfnamefont {R.~A.}\ \bibnamefont
  {Erickson}},\ }\bibfield  {title} {\bibinfo {title} {Obtaining and applying
  public data for training students in technical statistical writing: Case
  studies with data from us geological survey and general ecological
  literature},\ }\href@noop {} {\bibfield  {journal} {\bibinfo  {journal}
  {Journal of Statistics and Data Science Education}\ ,\ \bibinfo {pages} {1}}
  (\bibinfo {year} {2023})}\BibitemShut {NoStop}%
\bibitem [{\citenamefont {{Mathpix, Inc.}}(2020)}]{mathpix}%
  \BibitemOpen
  \bibfield  {author} {\bibinfo {author} {\bibnamefont {{Mathpix, Inc.}}},\
  }\href {https://mathpix.com} {\bibinfo {title} {Mathpix}} (\bibinfo {year}
  {2020})\BibitemShut {NoStop}%
\bibitem [{\citenamefont {Costa}\ \emph {et~al.}(2021)\citenamefont {Costa},
  \citenamefont {Mello},\ and\ \citenamefont {d'Amorim}}]{mathpix-benchmark}%
  \BibitemOpen
  \bibfield  {author} {\bibinfo {author} {\bibfnamefont {D.~S.}\ \bibnamefont
  {Costa}}, \bibinfo {author} {\bibfnamefont {C.~A.}\ \bibnamefont {Mello}},\
  and\ \bibinfo {author} {\bibfnamefont {M.}~\bibnamefont {d'Amorim}},\
  }\bibfield  {title} {\bibinfo {title} {A comparative study on methods and
  tools for handwritten mathematical expression recognition},\ }in\ \href@noop
  {} {\emph {\bibinfo {booktitle} {Proceedings of the 21st ACM Symposium on
  Document Engineering}}}\ (\bibinfo {year} {2021})\ pp.\ \bibinfo {pages}
  {1--4}\BibitemShut {NoStop}%
\bibitem [{\citenamefont {MacFarlane}()}]{pandoc}%
  \BibitemOpen
  \bibfield  {author} {\bibinfo {author} {\bibfnamefont {J.}~\bibnamefont
  {MacFarlane}},\ }\href {https://pandoc.org} {\bibinfo {title}
  {Pandoc}}\BibitemShut {NoStop}%
\bibitem [{\citenamefont {Kohlsch{\"u}tter}\ \emph {et~al.}(2010)\citenamefont
  {Kohlsch{\"u}tter}, \citenamefont {Fankhauser},\ and\ \citenamefont
  {Nejdl}}]{boilerplate}%
  \BibitemOpen
  \bibfield  {author} {\bibinfo {author} {\bibfnamefont {C.}~\bibnamefont
  {Kohlsch{\"u}tter}}, \bibinfo {author} {\bibfnamefont {P.}~\bibnamefont
  {Fankhauser}},\ and\ \bibinfo {author} {\bibfnamefont {W.}~\bibnamefont
  {Nejdl}},\ }\bibfield  {title} {\bibinfo {title} {Boilerplate detection using
  shallow text features},\ }in\ \href@noop {} {\emph {\bibinfo {booktitle}
  {Proceedings of the third ACM international conference on Web search and data
  mining}}}\ (\bibinfo {year} {2010})\ pp.\ \bibinfo {pages}
  {441--450}\BibitemShut {NoStop}%
\bibitem [{\citenamefont {Thoppilan}\ \emph {et~al.}(2022)\citenamefont
  {Thoppilan}, \citenamefont {De~Freitas}, \citenamefont {Hall}, \citenamefont
  {Shazeer}, \citenamefont {Kulshreshtha}, \citenamefont {Cheng}, \citenamefont
  {Jin}, \citenamefont {Bos}, \citenamefont {Baker}, \citenamefont {Du} \emph
  {et~al.}}]{lambda}%
  \BibitemOpen
  \bibfield  {author} {\bibinfo {author} {\bibfnamefont {R.}~\bibnamefont
  {Thoppilan}}, \bibinfo {author} {\bibfnamefont {D.}~\bibnamefont
  {De~Freitas}}, \bibinfo {author} {\bibfnamefont {J.}~\bibnamefont {Hall}},
  \bibinfo {author} {\bibfnamefont {N.}~\bibnamefont {Shazeer}}, \bibinfo
  {author} {\bibfnamefont {A.}~\bibnamefont {Kulshreshtha}}, \bibinfo {author}
  {\bibfnamefont {H.-T.}\ \bibnamefont {Cheng}}, \bibinfo {author}
  {\bibfnamefont {A.}~\bibnamefont {Jin}}, \bibinfo {author} {\bibfnamefont
  {T.}~\bibnamefont {Bos}}, \bibinfo {author} {\bibfnamefont {L.}~\bibnamefont
  {Baker}}, \bibinfo {author} {\bibfnamefont {Y.}~\bibnamefont {Du}}, \emph
  {et~al.},\ }\bibfield  {title} {\bibinfo {title} {Lamda: Language models for
  dialog applications},\ }\href@noop {} {\bibfield  {journal} {\bibinfo
  {journal} {arXiv preprint arXiv:2201.08239}\ } (\bibinfo {year}
  {2022})}\BibitemShut {NoStop}%
\bibitem [{\citenamefont {Anil}\ \emph {et~al.}(2023)\citenamefont {Anil},
  \citenamefont {Dai}, \citenamefont {Firat}, \citenamefont {Johnson},
  \citenamefont {Lepikhin}, \citenamefont {Passos}, \citenamefont {Shakeri},
  \citenamefont {Taropa}, \citenamefont {Bailey}, \citenamefont {Chen} \emph
  {et~al.}}]{palm}%
  \BibitemOpen
  \bibfield  {author} {\bibinfo {author} {\bibfnamefont {R.}~\bibnamefont
  {Anil}}, \bibinfo {author} {\bibfnamefont {A.~M.}\ \bibnamefont {Dai}},
  \bibinfo {author} {\bibfnamefont {O.}~\bibnamefont {Firat}}, \bibinfo
  {author} {\bibfnamefont {M.}~\bibnamefont {Johnson}}, \bibinfo {author}
  {\bibfnamefont {D.}~\bibnamefont {Lepikhin}}, \bibinfo {author}
  {\bibfnamefont {A.}~\bibnamefont {Passos}}, \bibinfo {author} {\bibfnamefont
  {S.}~\bibnamefont {Shakeri}}, \bibinfo {author} {\bibfnamefont
  {E.}~\bibnamefont {Taropa}}, \bibinfo {author} {\bibfnamefont
  {P.}~\bibnamefont {Bailey}}, \bibinfo {author} {\bibfnamefont
  {Z.}~\bibnamefont {Chen}}, \emph {et~al.},\ }\bibfield  {title} {\bibinfo
  {title} {Palm 2 technical report},\ }\href@noop {} {\bibfield  {journal}
  {\bibinfo  {journal} {arXiv preprint arXiv:2305.10403}\ } (\bibinfo {year}
  {2023})}\BibitemShut {NoStop}%
\bibitem [{\citenamefont {Bubeck}\ \emph {et~al.}(2023)\citenamefont {Bubeck},
  \citenamefont {Chandrasekaran}, \citenamefont {Eldan}, \citenamefont
  {Gehrke}, \citenamefont {Horvitz}, \citenamefont {Kamar}, \citenamefont
  {Lee}, \citenamefont {Lee}, \citenamefont {Li}, \citenamefont {Lundberg}
  \emph {et~al.}}]{gpt-stem}%
  \BibitemOpen
  \bibfield  {author} {\bibinfo {author} {\bibfnamefont {S.}~\bibnamefont
  {Bubeck}}, \bibinfo {author} {\bibfnamefont {V.}~\bibnamefont
  {Chandrasekaran}}, \bibinfo {author} {\bibfnamefont {R.}~\bibnamefont
  {Eldan}}, \bibinfo {author} {\bibfnamefont {J.}~\bibnamefont {Gehrke}},
  \bibinfo {author} {\bibfnamefont {E.}~\bibnamefont {Horvitz}}, \bibinfo
  {author} {\bibfnamefont {E.}~\bibnamefont {Kamar}}, \bibinfo {author}
  {\bibfnamefont {P.}~\bibnamefont {Lee}}, \bibinfo {author} {\bibfnamefont
  {Y.~T.}\ \bibnamefont {Lee}}, \bibinfo {author} {\bibfnamefont
  {Y.}~\bibnamefont {Li}}, \bibinfo {author} {\bibfnamefont {S.}~\bibnamefont
  {Lundberg}}, \emph {et~al.},\ }\bibfield  {title} {\bibinfo {title} {Sparks
  of artificial general intelligence: Early experiments with gpt-4},\
  }\href@noop {} {\bibfield  {journal} {\bibinfo  {journal} {arXiv preprint
  arXiv:2303.12712}\ } (\bibinfo {year} {2023})}\BibitemShut {NoStop}%
\bibitem [{\citenamefont {Frieder}\ \emph {et~al.}(2023)\citenamefont
  {Frieder}, \citenamefont {Pinchetti}, \citenamefont {Griffiths},
  \citenamefont {Salvatori}, \citenamefont {Lukasiewicz}, \citenamefont
  {Petersen}, \citenamefont {Chevalier},\ and\ \citenamefont
  {Berner}}]{gpt-math}%
  \BibitemOpen
  \bibfield  {author} {\bibinfo {author} {\bibfnamefont {S.}~\bibnamefont
  {Frieder}}, \bibinfo {author} {\bibfnamefont {L.}~\bibnamefont {Pinchetti}},
  \bibinfo {author} {\bibfnamefont {R.-R.}\ \bibnamefont {Griffiths}}, \bibinfo
  {author} {\bibfnamefont {T.}~\bibnamefont {Salvatori}}, \bibinfo {author}
  {\bibfnamefont {T.}~\bibnamefont {Lukasiewicz}}, \bibinfo {author}
  {\bibfnamefont {P.~C.}\ \bibnamefont {Petersen}}, \bibinfo {author}
  {\bibfnamefont {A.}~\bibnamefont {Chevalier}},\ and\ \bibinfo {author}
  {\bibfnamefont {J.}~\bibnamefont {Berner}},\ }\bibfield  {title} {\bibinfo
  {title} {Mathematical capabilities of chatgpt},\ }\href@noop {} {\bibfield
  {journal} {\bibinfo  {journal} {arXiv preprint arXiv:2301.13867}\ } (\bibinfo
  {year} {2023})}\BibitemShut {NoStop}%
\bibitem [{\citenamefont {Parr}\ and\ \citenamefont {Quong}(1995)}]{antlr}%
  \BibitemOpen
  \bibfield  {author} {\bibinfo {author} {\bibfnamefont {T.~J.}\ \bibnamefont
  {Parr}}\ and\ \bibinfo {author} {\bibfnamefont {R.~W.}\ \bibnamefont
  {Quong}},\ }\bibfield  {title} {\bibinfo {title} {Antlr: A predicated-ll (k)
  parser generator},\ }\href@noop {} {\bibfield  {journal} {\bibinfo  {journal}
  {Software: Practice and Experience}\ }\textbf {\bibinfo {volume} {25}},\
  \bibinfo {pages} {789} (\bibinfo {year} {1995})}\BibitemShut {NoStop}%
\bibitem [{\citenamefont {Larsen}\ \emph {et~al.}(2017)\citenamefont {Larsen},
  \citenamefont {Mortensen}, \citenamefont {Blomqvist}, \citenamefont
  {Castelli}, \citenamefont {Christensen}, \citenamefont {Du{\l}ak},
  \citenamefont {Friis}, \citenamefont {Groves}, \citenamefont {Hammer},
  \citenamefont {Hargus} \emph {et~al.}}]{ase}%
  \BibitemOpen
  \bibfield  {author} {\bibinfo {author} {\bibfnamefont {A.~H.}\ \bibnamefont
  {Larsen}}, \bibinfo {author} {\bibfnamefont {J.~J.}\ \bibnamefont
  {Mortensen}}, \bibinfo {author} {\bibfnamefont {J.}~\bibnamefont
  {Blomqvist}}, \bibinfo {author} {\bibfnamefont {I.~E.}\ \bibnamefont
  {Castelli}}, \bibinfo {author} {\bibfnamefont {R.}~\bibnamefont
  {Christensen}}, \bibinfo {author} {\bibfnamefont {M.}~\bibnamefont
  {Du{\l}ak}}, \bibinfo {author} {\bibfnamefont {J.}~\bibnamefont {Friis}},
  \bibinfo {author} {\bibfnamefont {M.~N.}\ \bibnamefont {Groves}}, \bibinfo
  {author} {\bibfnamefont {B.}~\bibnamefont {Hammer}}, \bibinfo {author}
  {\bibfnamefont {C.}~\bibnamefont {Hargus}}, \emph {et~al.},\ }\bibfield
  {title} {\bibinfo {title} {The atomic simulation environment—a python
  library for working with atoms},\ }\href@noop {} {\bibfield  {journal}
  {\bibinfo  {journal} {Journal of Physics: Condensed Matter}\ }\textbf
  {\bibinfo {volume} {29}},\ \bibinfo {pages} {273002} (\bibinfo {year}
  {2017})}\BibitemShut {NoStop}%
\bibitem [{\citenamefont {Ong}\ \emph {et~al.}(2013)\citenamefont {Ong},
  \citenamefont {Richards}, \citenamefont {Jain}, \citenamefont {Hautier},
  \citenamefont {Kocher}, \citenamefont {Cholia}, \citenamefont {Gunter},
  \citenamefont {Chevrier}, \citenamefont {Persson},\ and\ \citenamefont
  {Ceder}}]{pymatgen}%
  \BibitemOpen
  \bibfield  {author} {\bibinfo {author} {\bibfnamefont {S.~P.}\ \bibnamefont
  {Ong}}, \bibinfo {author} {\bibfnamefont {W.~D.}\ \bibnamefont {Richards}},
  \bibinfo {author} {\bibfnamefont {A.}~\bibnamefont {Jain}}, \bibinfo {author}
  {\bibfnamefont {G.}~\bibnamefont {Hautier}}, \bibinfo {author} {\bibfnamefont
  {M.}~\bibnamefont {Kocher}}, \bibinfo {author} {\bibfnamefont
  {S.}~\bibnamefont {Cholia}}, \bibinfo {author} {\bibfnamefont
  {D.}~\bibnamefont {Gunter}}, \bibinfo {author} {\bibfnamefont {V.~L.}\
  \bibnamefont {Chevrier}}, \bibinfo {author} {\bibfnamefont {K.~A.}\
  \bibnamefont {Persson}},\ and\ \bibinfo {author} {\bibfnamefont
  {G.}~\bibnamefont {Ceder}},\ }\bibfield  {title} {\bibinfo {title} {Python
  materials genomics (pymatgen): A robust, open-source python library for
  materials analysis},\ }\href@noop {} {\bibfield  {journal} {\bibinfo
  {journal} {Computational Materials Science}\ }\textbf {\bibinfo {volume}
  {68}},\ \bibinfo {pages} {314} (\bibinfo {year} {2013})}\BibitemShut
  {NoStop}%
\bibitem [{\citenamefont {Togo}\ and\ \citenamefont {Tanaka}(2015)}]{phonopy}%
  \BibitemOpen
  \bibfield  {author} {\bibinfo {author} {\bibfnamefont {A.}~\bibnamefont
  {Togo}}\ and\ \bibinfo {author} {\bibfnamefont {I.}~\bibnamefont {Tanaka}},\
  }\bibfield  {title} {\bibinfo {title} {First principles phonon calculations
  in materials science},\ }\href@noop {} {\bibfield  {journal} {\bibinfo
  {journal} {Scr. Mater.}\ }\textbf {\bibinfo {volume} {108}},\ \bibinfo
  {pages} {1} (\bibinfo {year} {2015})}\BibitemShut {NoStop}%
\bibitem [{\citenamefont {Behler}\ and\ \citenamefont
  {Parrinello}(2007)}]{behler-parrinello}%
  \BibitemOpen
  \bibfield  {author} {\bibinfo {author} {\bibfnamefont {J.}~\bibnamefont
  {Behler}}\ and\ \bibinfo {author} {\bibfnamefont {M.}~\bibnamefont
  {Parrinello}},\ }\bibfield  {title} {\bibinfo {title} {Generalized
  neural-network representation of high-dimensional potential-energy
  surfaces},\ }\href@noop {} {\bibfield  {journal} {\bibinfo  {journal}
  {Physical review letters}\ }\textbf {\bibinfo {volume} {98}},\ \bibinfo
  {pages} {146401} (\bibinfo {year} {2007})}\BibitemShut {NoStop}%
\bibitem [{\citenamefont {Paszke}\ \emph {et~al.}(2019)\citenamefont {Paszke},
  \citenamefont {Gross}, \citenamefont {Massa}, \citenamefont {Lerer},
  \citenamefont {Bradbury}, \citenamefont {Chanan}, \citenamefont {Killeen},
  \citenamefont {Lin}, \citenamefont {Gimelshein}, \citenamefont {Antiga} \emph
  {et~al.}}]{pytorch}%
  \BibitemOpen
  \bibfield  {author} {\bibinfo {author} {\bibfnamefont {A.}~\bibnamefont
  {Paszke}}, \bibinfo {author} {\bibfnamefont {S.}~\bibnamefont {Gross}},
  \bibinfo {author} {\bibfnamefont {F.}~\bibnamefont {Massa}}, \bibinfo
  {author} {\bibfnamefont {A.}~\bibnamefont {Lerer}}, \bibinfo {author}
  {\bibfnamefont {J.}~\bibnamefont {Bradbury}}, \bibinfo {author}
  {\bibfnamefont {G.}~\bibnamefont {Chanan}}, \bibinfo {author} {\bibfnamefont
  {T.}~\bibnamefont {Killeen}}, \bibinfo {author} {\bibfnamefont
  {Z.}~\bibnamefont {Lin}}, \bibinfo {author} {\bibfnamefont {N.}~\bibnamefont
  {Gimelshein}}, \bibinfo {author} {\bibfnamefont {L.}~\bibnamefont {Antiga}},
  \emph {et~al.},\ }\bibfield  {title} {\bibinfo {title} {Pytorch: An
  imperative style, high-performance deep learning library},\ }\href@noop {}
  {\bibfield  {journal} {\bibinfo  {journal} {Advances in neural information
  processing systems}\ }\textbf {\bibinfo {volume} {32}} (\bibinfo {year}
  {2019})}\BibitemShut {NoStop}%
\bibitem [{\citenamefont {Kingma}\ and\ \citenamefont {Ba}(2014)}]{adam}%
  \BibitemOpen
  \bibfield  {author} {\bibinfo {author} {\bibfnamefont {D.~P.}\ \bibnamefont
  {Kingma}}\ and\ \bibinfo {author} {\bibfnamefont {J.}~\bibnamefont {Ba}},\
  }\bibfield  {title} {\bibinfo {title} {Adam: A method for stochastic
  optimization},\ }\href@noop {} {\bibfield  {journal} {\bibinfo  {journal}
  {arXiv preprint arXiv:1412.6980}\ } (\bibinfo {year} {2014})}\BibitemShut
  {NoStop}%
\bibitem [{\citenamefont {McInnes}\ \emph {et~al.}(2018)\citenamefont
  {McInnes}, \citenamefont {Healy},\ and\ \citenamefont {Melville}}]{umap}%
  \BibitemOpen
  \bibfield  {author} {\bibinfo {author} {\bibfnamefont {L.}~\bibnamefont
  {McInnes}}, \bibinfo {author} {\bibfnamefont {J.}~\bibnamefont {Healy}},\
  and\ \bibinfo {author} {\bibfnamefont {J.}~\bibnamefont {Melville}},\
  }\bibfield  {title} {\bibinfo {title} {Umap: Uniform manifold approximation
  and projection for dimension reduction},\ }\href@noop {} {\bibfield
  {journal} {\bibinfo  {journal} {arXiv preprint arXiv:1802.03426}\ } (\bibinfo
  {year} {2018})}\BibitemShut {NoStop}%
\end{thebibliography}%

\end{document}